\documentclass[11pt]{article}
\usepackage{graphicx}

\begin{document} 

\def\xslash#1{{\rlap{$#1$}/}}
\def \p {\partial}
\def \dd {\psi_{u\bar dg}}
\def \ddp {\psi_{u\bar dgg}}
\def \pq {\psi_{u\bar d\bar uu}}
\def \jpsi {J/\psi}
\def \psip {\psi^\prime}
\def \txi2 {\tilde\xi_2} 
\def \to {\rightarrow}
\def \lrto{\leftrightarrow} 
\def\bfsig{\mbox{\boldmath$\sigma$}}
\def\DT{\mbox{\boldmath$\Delta_T $}}
\def\xit{\mbox{\boldmath$\xi_\perp $}}
\def \jpsi {J/\psi}
\def\bfej{\mbox{\boldmath$\varepsilon$}}
\def \t {\tilde}
\def\epn {\varepsilon}
\def \up {\uparrow}
\def \dn {\downarrow}
\def \da {\dagger}
\def \pn3 {\phi_{u\bar d g}}

\def \p4n {\phi_{u\bar d gg}}

\def \bx {\bar x}
\def \by {\bar y}



\begin{center}
{\Large\bf One-Loop Corrections of Single Spin Asymmetries in Semi-Inclusive DIS }
\par\vskip20pt
A.P. Chen$^{1,2,3}$, J.P. Ma$^{1,2,3}$, G.P. Zhang$^{4}$     \\
{\small {\it
$^1$ Institute of Theoretical Physics, Chinese Academy of Sciences,
P.O. Box 2735,
Beijing 100190, China\\
$^2$ School of Physical Sciences, University of Chinese Academy of Sciences, Beijing 100049, China\\
$^3$ School of Physics and Center for High-Energy Physics, Peking University, Beijing 100871, China\\
$^4$Department of Modern Physics,  University of Science and Technology of China, Hefei, Anhui 230026, China  
}} \\
\end{center}
\vskip 1cm
\begin{abstract}
We study single spin asymmetries at one-loop accuracy in semi-inclusive DIS with a transversely polarized hadron in the initial state. 
Two measurable spin observables are predicted in the framework of QCD collinear factorization.
One of the spin observables is the Sivers weighted asymmetry, another one is the Collins weighted asymmetry.   
The prediction takes a form of convolutions of perturbative coefficient functions and nonperturbative functions, which are twist-2 transversity distributions, twist-3 parton distributions, twist-2- and twist-3 parton fragmentation functions. 
These nonperturbative functions can be extracted from measurements of the spin observables and provide 
valuable information of the inner structure of hadrons. The measurements can be done in current COMPASS- and JLab experiments
and in future experiments of EIC. The perturbative coefficient functions are calculated at one-loop level.
There are collinear divergences in the calculation involving chirality-even- and chirality-odd twist operators. We find 
that all collinear divergences can be correctly subtracted so that the final results are finite.

\vskip 5mm
\noindent
\end{abstract}
\vskip 1cm

\par\vskip20pt
\noindent 
{\bf 1. Introduction}
\par
Experiments of lepton-hadron scattering have provided important information about the inner structure 
of hadrons. A typical example is Deeply Inelastic Scattering(DIS). From DIS one can extract parton distribution functions 
defined with twist-2 operators of QCD.  
In Semi-Inclusive DIS(SIDIS) with one detected hadron in the final state, one can learn more about the inner structure and 
nonperturbative properties of QCD, if the initial hadron is transversely polarized. In this case the spin-dependent 
part of the differential cross-section, or Single-Spin Asymmetries(SSA's), can be predicted  with parton distributions 
defined with twist-3 operators as shown in \cite{EFTE, QiuSt}, and twist-3 Fragmentation Functions(FF's). The twist-3 parton distributions describe quark-gluon correlations inside a hadron and  contain more information about the inner structure of hadrons than twist-2 parton distributions. In this work, we study SSA's in SIDIS, in particular one-loop corrections of spin-observables.      

\par 
Under one-photon-exchange approximation, the hadronic tensor of SIDIS contains all information about the process. 
The spin-dependent part of the tensor has been studied with collinear factorization at tree-level in different kinematic regions. In the region 
where the final hadron has large transverse momentum $P_{h\perp}$, SSA has been studied in \cite{EKT,EKT2,KaKo}, where 
the spin-dependent part starts at order of $\alpha_s$. At low $P_{h\perp}$ one can also employ collinear factorization
because of the large virtuality $-Q^2$ of the exchanged photon. In this region, the spin-dependent part starting at order 
of $\alpha_s^0$ is predicted 
as a tensor distribution of $P_{h\perp}$ as shown in \cite{CMZ}. This implies that the measurable effects from this part  can only be 
predicted when $P_{h\perp}$ is integrated over with certain weights. In this work we construct two spin-observables by integrating over $P_{h\perp}$ with different weights, and study one-loop corrections of the two spin-observables.

\par 
Various SSA's in SIDIS can be measured in current experiment of COMPASS\cite{COMP} and JLab\cite{JLab1} 
and in future experiment of EIC\cite{EIC1}. It is noted in the low $P_{h\perp}$ region one can employ the approach of Transverse-Momentum-Dependent(TMD) factorization studied in \cite{CS,CSS,JMY}, where nonperturbative effects are described
by TMD parton distributions and TMD FF's. The relevant phenomenology with TMD factorization for SIDIS has been studied 
 in \cite{PMRT,DBPM,DBJPM,NSIDIS}.  In the framework of TMD factorization,  the SSA related to Sivers function,  which is one of TMD parton distributions, is called Sivers Asymmetry. The SSA  related to Collins fragmentation function 
 is called Collins asymmetry. These two asymmetries have already been studied in experiment by HERMES and COMPASS. It is found that these asymmetries are different than zero\cite{Hermes,CompA}. Recently, the so-called 
 weighted Sivers asymmetry has been studied by COMPASS\cite{CompAR}.  One of our two spin observables
 is in fact the studied weighted Sivers asymmetry. Another one is the weighted Collins asymmetry, because it is related to the transverse momentum moment of Collins fragmentation function.  These two asymmetries have important implications, if they are not zero.  If Sivers asymmetry is not zero, it indicates that partons inside a hadron have nonzero 
 orbital angular momentum.  Nonzero Collins asymmetry indicates that partons in their fragmentation into a hadron 
 can have nonzero orbital angular momentum.

\par
Without observing the spin of the final hadron, twist-3 contributions only appear in the case when the initial hadron is transversely polarized. The spin-dependent part of the hadronic tensor can obtain contributions from twist-3 parton distributions introduced in \cite{EFTE, QiuSt} combined with twist-2 quark FF. 
These twist-3 parton distributions and twist-2 FF are defined with chirality-even operators. We will call 
these contributions as chirality-even contributions. Besides them the spin-dependent part also receives contributions   
from  the twist-2 transversity distribution introduced in \cite{JaJi}, combined with twist-3 FF's.
The transversity distribution and twist-3 FF's are defined with chirality-odd operators. We will 
call these contributions as chirality-odd contributions. At the leading order of $\alpha_s$, i.e., at $\alpha_s^0$, one of our two spin observables has only a chirality-even contributions, while another 
has only a chirality-odd contribution. Therefore, through the two observables one can extract not only twist-3 parton distributions and fragmentation functions, but also the twist-2 transversity distribution, which is less known than 
other twist-2 parton distributions. Beyond tree-level, each spin observable can have chirality-even- and chirality-odd 
contributions.        

\par 
It is worth to point out that there are not many results of one-loop calculation involving twist-3 operators, while 
calculations beyond tree-level with only twist-2 operators are rather standard and many one-loop results exist. 
For Drell-Yan processes with one transversely polarized hadron in the initial state, one-loop correction of a spin observable 
involving the twist-3 parton distributions is calculated at one-loop in \cite{VoYu}. In \cite{CMZDY} 
two spin observables are studied and their complete one-loop corrections are derived.  For SIDIS, different parts of one-loop chirality-even correction      
have been studied in \cite{KVX,DKPV,ShYo} for one of our two spin observables. But the one-loop chirality-even corrections from \cite{KVX,DKPV,ShYo}
are still not completed and the one-loop chirality-odd corrections are missing. In this work we will give complete one-loop 
corrections of the two spin observables. One-loop study of twist-3 effect for DIS has been performed in \cite{G2}.

\par 
An interesting observation has been made for the twist-3 part of the hadronic tensor in \cite{CMZ}.
The twist-3 part at tree-level has contributions proportional to the derivative of $\delta^2(P_{h\perp})$. The virtual corrections 
beyond tree-level of these contributions are completely determined by the loop corrections of the quark form factor.   
Similar observation has been also made for Drell-Yan processes in \cite{MaZh3}. In this work, the two spin observables 
are so constructed that they receive at tree-level only from those contributions with the derivative of $\delta^2(P_{h\perp})$ of the twist-3 parts. Then the virtual correction of the spin observables can be obtained from the relevant results of the 
quark form factor, we will mainly deal with the real correction.

\par
Calculations involving twist-3 operators are in general more complicated than those of twist-2. The nonperturbative- and 
perturbative effects must be separated in a gauge-invariant way. This has been discussed in detail in \cite{EKT2} for 
SIDIS. Unlike the twist-2 factorization, where partons can never have zero momentum fraction, 
in the twist-3 factorization some partons participating hard scattering can have zero momentum fraction. 
In \cite{QiuSt} it has been shown there are  so-called soft-gluon-pole contributions, in which one gluon as a parton 
has zero momentum. The gluon is not exactly with zero momentum. In fact, as shown in \cite{MSS},  
it is a Glauber gluon and its momentum can be neglected in hard scattering. It is difficult to calculate the soft-gluon-pole contributions. However, these contributions at tree-level can be related to the corresponding twist-2 contribution, as shown 
in \cite{KoTa,KoTa2,KTY}. There are so-called master formulas to obtain the contributions. This will simplify 
our calculation of one-loop real correction, since it is a tree-level calculation before that some final states 
are summed. With the results for SIDIS in \cite{EKT2,KoTa}, the twist-3 calculations of SIDIS can be performed 
straightforwardly.

\par

Our paper is organized as follows. In Sect. 2 we introduce our notations. We define two spin observables and derive the tree-level results. In Sect. 3 and Sect. 4
we give the one-loop corrections for the chirality-odd- and chirality-even contributions, respectively. In these sections, 
we also perform the subtraction of the collinear contributions. The collinear singularities will be subtracted into 
various parton distributions and FF's. The final results are finite. 
In Sect. 5 we give our final results. Sect. 6 is our summary. In Appendix we list perturbative coefficient functions 
of our one-loop corrections.

\par\vskip20pt
\noindent
{\bf 2. Notations and Tree-Level Results}
\par
We consider the semi-inclusive process:
\begin{equation} 
   e(k) + h(P,s) \to e(k') + h'(P_h) + X
\end{equation} 
where the initial hadron $h$ is a spin-1/2 one with the spin vector $s$. We will consider 
the case that the polarization of particles in the final state is not observed or 
summed over and the initial electron is unpolarized. 
At the leading order of QED,  
the process is described by the hadronic tensor: 
\begin{equation} 
W^{\mu\nu} = \sum_X \int \frac {d^4 x}{(2\pi)^4} e^{iq\cdot x} \langle P,s\vert J^\mu (x) \vert P_h, X\rangle 
     \langle P_h, X\vert J^\nu (0) \vert P, s\rangle 
\end{equation} 
with $q=k-k'$ as the momentum of the virtual photon emitted from the initial electron.
We will consider the process in the kinematic region with $Q^2=-q^2\gg \Lambda^2_{QCD}$. In this region one can use the concept 
of QCD factorization to predict $W^{\mu\nu}$ in the form of convolutions with perturbative coefficient functions, 
various parton distributions and FF's. We are interested in the transverse-spin-dependent part of $W^{\mu\nu}$.      
In this part twist-3 parton distributions and twist-3 FF's are involved.
\par 
To define parton distributions and FF's it is convenient to use the light-cone coordinate system. 
In this system a
vector $a^\mu$ is expressed as $a^\mu = (a^+, a^-, \vec a_\perp) =
((a^0+a^3)/\sqrt{2}, (a^0-a^3)/\sqrt{2}, a^1, a^2)$ and $\vec a_\perp^2
=(a^1)^2+(a^2)^2=-a_\perp\cdot a_\perp$. We introduce two light-cone vectors as $l^\mu=(1,0,0,0)$ and $n^\mu =(0,1,0,0)$. 
With these two vectors one can define: 
\begin{equation}
  g_\perp^{\mu\nu} = g^{\mu\nu} - n^\mu l^\nu - n^\nu l^\mu,
  \quad
  \epsilon_\perp^{\mu\nu} =\epsilon^{\alpha\beta\mu\nu}l_\alpha n_\beta. 
\label{2TT}   
\end{equation}
We take the initial hadron moving in the $z$-direction with the momentum $P^\mu=(P^+,0,0,0)$. The initial hadron is transversely 
polarized with $s^\mu=(0,0,s_\perp^1,s_\perp^2)$. At twist-2 there is one parton distribution related to the 
transverse spin. It is the transversity distribution introduced first in \cite{JaJi}:  
\begin{equation} 
 \int \frac{ d\lambda }{4\pi} e^{-ix \lambda P^+} \langle P,s_\perp \vert  \bar \psi_i (\lambda n)   
    \psi_j (0)   \vert P,s_\perp\rangle = \frac{1}{4 N_c P^+ } \biggr ( \gamma_\perp\cdot s_\perp \gamma\cdot P  h_1(x) +\cdots 
      \biggr )_{ji}, 
\label{LCH1}            
\end{equation}
where $ij$ stand for Dirac indices and color indices and $\cdots$ denote irrelevant terms. Here and in the following we suppress gauge links between field operators for brevity.   
$x$ is the momentum fraction carried by the quark.  This distribution is defined with the operator which is chirality-odd. Hence, the contributions
to $W^{\mu\nu}$ involving $h_1$ will always be combined with chirality-odd FF's.    

\par 
At twist-3 there are two transverse-spin dependent parton distributions defined with quark-gluon-quark correlations. 
They are the so-called ETQS matrix elements defined in \cite{EFTE,QiuSt}:  
\begin{eqnarray}
&& \int \frac{d \lambda_1 d\lambda_2}{4\pi}
e^{ -i\lambda_2 (x_2-x_1) P^+ -i \lambda_1 x_1 P^+  }
 \langle P, s_\perp  \vert
           \bar\psi_i (\lambda_1n ) g_s G^{+\mu}(\lambda_2n) \psi_j(0) \vert P,s_\perp \rangle
\nonumber\\
  && = \frac{1}{4} \left [ \gamma^ - \right ]_{ji} \tilde s_{\perp}^{\mu} T_F(x_1,x_2)
    + \frac{1}{4} \left [ i \gamma_5 \gamma^- \right ]_{ji} s^\mu_\perp T_\Delta (x_1,x_2) +\cdots,
\label{tw3q}     
\end{eqnarray}
where $\tilde s_\perp^\mu$ is defined as $\tilde s_\perp^\mu = \epsilon_\perp^{\mu\nu} s_{\perp\nu}$ and $\cdots$ denote irrelevant terms.   The two twist-3 parton distribution functions have the property:
\begin{equation} 
  T_F (x_1,x_2) = T_F(x_2,x_1),\quad T_\Delta (x_1,x_2) = -T_\Delta (x_2,x_1). 
\end{equation}   
Replacing the field-strength tensor 
operator in Eq.(\ref{tw3q}) with the covariant derivative $D_\perp^\mu$ one can define other two twist-3 distributions. 
There are three twist-3 distributions defined with a product of two quark field operators. Two of them 
are given in \cite{EKT}, and one of them is defined in \cite{CMZ}. These twist-3 distributions can be expressed with the two defined in Eq.(\ref{tw3q})\cite{EKT,CMZ}.   
\par 
Four twist-3 distributions can be defined with purely gluonic operators\cite{Ji3G}. One of them can be defined as:   
\begin{eqnarray} 
T^{(f)}_{G} (x_1,x_2) \tilde s^\mu  &=&   \frac{ i g_s f^{abc} g_{\perp\alpha\beta} }{P^+} 
      \int\frac{d y_1 d y_2}{4\pi} 
   e^{-i P^+ (y_2 (x_2-x_1) + y_1 x_1)}
\nonumber\\   
    && \langle P, s_\perp\vert G^{a,+\alpha}(y_1 n) G^{b, +\mu}(y_2 n) G^{c,+\beta} (0) \vert P, s_\perp \rangle, 
\label{3GT3}      
\end{eqnarray}
Replacing $if^{abc}$ 
with $d^{abc}$ one obtains the definition of  $T^{(d)}_{G}$. Besides these two distributions $T^{(f,d)}_{G}$ other two twist-3 distributions are defined by replacing $g_{\perp\alpha\beta}$ 
with $\epsilon_{\perp\alpha\beta}$ in Eq.(\ref{3GT3}). But, the contributions with these two twist-3 distributions 
do not appear in the two spin observables studied in this work. 
For the matrix elements
with $f^{abc}$  one has:
\begin{eqnarray}
T^{(f)}_{G}(x_1,x_2) = -T^{(f)}_G(-x_2,-x_1), \quad T^{(f)}_G(x_1,x_2) = T^{(f)}_G(x_2,x_1), 
\label{TSY} 
\end{eqnarray}
Similar relations can be derived for distributions defined with $d^{abc}$.

\par
To define FF's, we assume  that the produced hadron moves in the $-z$-direction with the momentum $P_h^\mu =(0,P_h^-,0,0)$. From two-parton correlations we define: 
\begin{eqnarray} 
  \Gamma_{ji} (k) &=& \int \frac{d^4 \xi}{(2\pi)^4} e^{-i \xi\cdot k} \sum_X \langle  0\vert \psi_j (0)  \vert P_h X \rangle  \langle  P_h X \vert \bar \psi_i (\xi) \vert 0 \rangle 
 \nonumber\\
      &=&    \frac{\delta (k^+) }{ z^{d-3} P^-_h} \biggr [ \delta^2 (k_\perp) \biggr ( \gamma^+ P_h^- \hat d(z)+  \hat e (z) +  \sigma^{-+}  \hat e_I(z)   \biggr )   - i \gamma^+ \gamma_\perp^\mu  P_h^- \hat e_\partial (z) \frac{\partial}{\partial k_\perp^\mu }\delta^2 (k_\perp)\biggr ]_{ji} 
   +\cdots, 
\label{2PFF}        
\end{eqnarray} 
where $ij$ stand for Dirac indices and color indices and $k^-$ is fixed as $P_h^-/z$. 
$\hat d(z)$ is the standard twist-2 FF\cite{PDFFF}.  $\hat e$, $\hat e_I$ and $\hat e_\partial$ are of twist-3.  $\hat e$, $\hat e_I$ have been first introduced in  \cite{JiFF}.  
From three-parton correlations one can define three twist-3 FF's. Two of them are: 
\begin{eqnarray} 
 \hat E_F (z_1,z_2) &=&  - \frac{z_2^{d-3} g_s}{2(d-2) N_c}  \int \frac{d\lambda_1 d\lambda_2}{(2\pi)^2} e^{-i\lambda_1 P_h^-/z_1 -i\lambda_2P_h^-/{z_3} }  
\nonumber\\ 
  && \sum_X {\rm Tr} \langle 0\vert i \gamma^- \gamma_{\perp\mu}  \psi(0) \vert P_h X\rangle \langle P_h X \vert \bar \psi (\lambda_1 l ) G^{-\mu}(\lambda_2 l) 
   \vert 0 \rangle ,
\nonumber\\
   \hat E_G (z_1,z_2) &=& -\frac{z_2^{d-3} g_s}{4 (N_c^2-1) } \frac{2}{d-2} \int \frac{d\lambda_1 d\lambda_2}{(2\pi)^2} e^{i \lambda_1 P_h^-/z_1 -i\lambda_2P_h^-/z_2  }
\nonumber\\  
   && \sum_X {\rm Tr} \langle 0\vert \bar \psi (\lambda_1 l ) i \gamma^- \gamma_{\perp\mu} T^a \psi(0) \vert P_h X\rangle \langle P_h X \vert  G^{a, -\mu}(\lambda_2 l) 
   \vert 0 \rangle
\label{3PFF}     
\end{eqnarray}
with $1/{z_3} = 1/z_2 -1/z_1$. Through charge conjugation of the operator in $\hat E_F$ one can define anther FF $\hat E_{\bar F}$ which is for fragmentation with an anti-quark.  
Similarly one can define an additional FF $\hat E_D$ by replacing 
$g_s G^{-\mu}(\lambda_2l)$ with $P_h^- D_\perp^\mu (\lambda_2 l)$. But this function 
is completely determined by $\hat E_F$ and $\hat e_\partial$\cite{MPF}. 
$\hat e(z)$ and $\hat e_{I,\partial}(z)$ have the support $\vert z\vert < 1$. For $z>0$, these FF's are for fragmentation 
of a quark. $\hat E_{F, G}(z_1,z_2)$ has the support\cite{MeMe}:
\begin{eqnarray} 
  0 < z_2 < 1,\quad {\rm or } \quad  z_2 < z_1 <\infty.  
\label{Z12R}    
\end{eqnarray}   
In \cite{MeMe} it is shown that these functions are zero at $z_1=z_2$ or  $1/z_1=0$, i.e., no parton in these FF's can have 
zero momentum fraction. 
\par 
All introduced 
twist-3 FF's are chirality-odd. The functions
$\hat e$, $\hat e_I$ and $\hat e_\partial$ are real, while $\hat E_{F, G}$ is complex in general.
If there are no final-state interactions, $\hat e_I$ and $\hat e_\partial$ are zero and $\hat E_{F,\bar F, G}$ is real.    
It is shown in \cite{MPF} that there are relations among these twist-3 FF's. In our notations they are: 
\begin{equation}
 z_2^2 \int \frac{d z_1}{z_1} P\frac{1}{ z_2-z_1} {\rm Im } \hat E_F (z_1,z_2) = 2 z_2 \hat e_\partial (z_2) - \hat e_I(z_2), 
 \quad \hat e(z_2) = z_2^2 \int \frac{d z_1}{z_1} P\frac{1}{z_2-z_1} {\rm Re }\hat E_F(z_1,z_2). 
\label{RL2}
\end{equation}  
In this work, we will take $\hat e_I$ and $\hat e$ as redundant in calculations of one-loop corrections. 
\par 
The standard variables for the considered process are:  
\begin{equation} 
x_B = \frac{Q^2}{2 P\cdot q},\quad  y=\frac{ P\cdot q}{P\cdot k}, \quad z_h=\frac{P\cdot P_h}{P\cdot q}.  
\end{equation} 
It is convenient to take the frame for the process in which the initial hadron moves in the $z$-direction, and 
the virtual photon moves in the $-z$-direction. In this frame the relevant momenta are given by:
\begin{equation} 
  q^\mu =(q^+, q^-,0,0), \quad P^\mu \approx (P^+,0,0,0), \quad P_h^\mu =(P_h^+, P_h^-, P_{h\perp}^1, P_{h\perp}^2), 
\quad P_{h\perp}^\mu =  g_\perp^{\mu\nu} P_{h\nu}.   
\end{equation}
To simplify notations, we will give our results for QCD with one flavor quark and its electric charge fraction  is set to be 1. 
It is easy to generalize our results to the case of multi-flavor quarks and to implement the real  electric charges. 
\par  
In \cite{CMZ} the transverse-spin dependent part of $W^{\mu\nu}$ at tree-level has been derived. The symmetric and twist-3 part of $W^{\mu\nu}$ is given by: 
\begin{eqnarray}
W^{\mu\nu}&=&  \frac{2} {x_B P\cdot q}\delta^2 (P_{h\perp})  
   \biggr ( (2 x_B P+q)^\mu \tilde s_\perp^\nu +  (2 x_B P+q)^\nu\tilde s_\perp^\mu \biggr )    h_1(x_B)  \biggr (  2 z_h \hat  e_\partial (z_h) -\hat e_I (z_h) \biggr )   
\nonumber\\
   && - z_h^2   \frac{\partial}{\partial P_{h\perp}^\rho} \delta^2(P_{h\perp} ) \biggr [   
      T_F(x_B,x_B) \hat d(z_h)   g_\perp^{\mu\nu} \tilde s_\perp^\rho 
  +2  \biggr ( g_\perp^{\mu\nu} \tilde s_\perp^\rho -g_\perp^{\mu\rho}\tilde s_\perp^\nu 
       -g_\perp^{\nu\rho}\tilde s_\perp^\mu \biggr ) h_1 (x_B) \hat e_\partial (z_h)  \biggr ].  
\label{TreeW}   
\end{eqnarray} 
This expression is explicitly $U(1)$-gauge invariant in the frame because of $q_\perp^\mu =0$.
Although the two transverse tensors $g_\perp^{\mu\nu}$ and $\epsilon_\perp^{\mu\nu}$ from Eq.(\ref{2TT}) are used here, 
the above result and $P_{h\perp}^\mu$ are covariant by realizing that the two transverse tensors can be defined covariantly:
\begin{equation}
g_\perp^{\mu\nu} = g^{\mu\nu} -\frac{1}{P\cdot \bar P} \left ( P^\mu \bar P^\nu 
 + P^\nu \bar P^\mu\right ),\quad \epsilon_\perp^{\mu\nu} =  \frac{1}{P\cdot \bar P} \epsilon^{\alpha\beta\mu\nu} P_\alpha \bar P_\beta , \quad \bar P = x_B P +q. 
\end{equation} 
Beyond tree-level, Eq.(\ref{TreeW}) receives corrections starting at order of $\alpha_s$. 
In the derivation of the result in \cite{CMZ} it is found that the virtual correction to the derivative part of hadronic tensors is determined by the correction 
of the quark form factor. Based on this result, the virtual correction to the second line in Eq.(\ref{TreeW}) 
is determined by the correction of the quark form factor. 

\par

The kinematics of the process has been discussed in detail in \cite{NSIDIS,Diehl}.  
The differential cross-section is given by: 
\begin{equation} 
   \frac{ d\sigma}{d x_B d y d z_h d\psi d^2  P_{h\perp} } = \frac{\alpha^2 y}{4 z_h Q^4} L_{\mu\nu} W^{\mu\nu}, 
\label{DIFFD}   
\end{equation} 
where $\psi$ is the azimuthal angle of the outgoing lepton.                 
$L^{\mu\nu}$ is the leptonic tensor: 
\begin{equation} 
L^{\mu\nu} = 2 \left ( k^\mu k'^{\nu} +  k^\nu k'^{\mu} -k\cdot k' g^{\mu\nu} \right ).  
\end{equation}  
In principle one can measure the differential cross-section in Eq.(\ref{DIFFD}) to detect the twist-3 effects or SSA's, 
because tat at twist-3 $W^{\mu\nu}$ is predicted as a tensor distribution, indicated by the $\delta$-function 
$\delta^2(P_{h\perp})$ and its derivative. This implies that measurable quantities can only be predicted  
when one integrates in Eq.(\ref{DIFFD}) over $P_{h\perp}$ with some weights depending on $P_{h\perp}$. We can define these
measurable quantities or  observables as weighted 
differential cross-sections as:   
\begin{equation} 
   \frac{ d\sigma \langle \mathcal O\rangle }{d x_B d y d z_h } = \frac{\alpha^2 y}{4 z_h Q^4} \int d\psi d^2 P_{h\perp} {\mathcal O}  L_{\mu\nu} W^{\mu\nu},
\label{IOS}     
\end{equation} 
with ${\mathcal O}$ as a weight depending on $P_h, k'$ and $s_\perp$. We will use dimensional regularization with 
$d=4-\epsilon$ to regularize divergences, Eq.(\ref{IOS}) should be understood as in $d$-dimensional space-time. 
In our final results one can set $d=4$ because they are finite.  
In this work we will study two spin-observables. They are defined 
with the weights: 
\begin{equation} 
  {\mathcal O}_1 = P_{h\perp}\cdot \tilde s_\perp, \quad {\mathcal O}_2 = P_{h\perp}\cdot k'_\perp k'_\perp\cdot \tilde s_\perp  - \frac{1}{2-\epsilon} \frac{Q^2 (y-1)}{y^2} P_{h\perp}\cdot \tilde s_\perp.   
\end{equation}
These two weights are proportional to $P_{h\perp}$. 
To clarify the meaning of these two observables, 
we take a frame in which the initial hadron moves in the $z$-direction and the virtual photon moves in the $-z$-direction. 
In this frame we denote the azimuthal angle between the lepton plane and the observed hadron $h'$ in the final state 
as $\phi_h$ and the azimuthal angle between the lepton plane and the transverse spin  
as $\phi_s$.  The two weights are related to these azimuthal angles as:
\begin{equation} 
{\mathcal O}_1 =- \vert P_{h\perp} \vert  \vert s_\perp\vert \sin(\phi_h-\phi_s), 
\quad {\mathcal O}_2 = -  Q^2 \frac{1-y}{2 y^2} \vert P_{h\perp} \vert \vert s_\perp\vert  \sin(\phi_h +\phi_s). 
\end{equation}    
It is noted that nonzero Sivers- or Collins asymmetry indicates that the azimuthal-angle distribution has a nonzero contribution 
proportional to $  \sin(\phi_h-\phi_s)$ or $\sin(\phi_h +\phi_s)$, respectively.  Therefore, the differential cross-section in Eq.(\ref{IOS}) with the weight ${\mathcal O}_1$ or ${\mathcal O}_2$  is proportional to the coefficient in the front 
of $\sin(\phi_h-\phi_s)$ or $\sin(\phi_h +\phi_s)$ in the azimuthal-angle distribution, respectively.  Hence, our two 
spin observables are  weighted Sivers- or Collins asymmetry.  It is noted that one can construct observables
more than two given in this work as shown, e.g., in \cite{CMZ}.

\par 

Substituting the two weights into Eq.(\ref{IOS}), it is easy to find that only the part of $W^{\mu\nu}$ in Eq.(\ref{TreeW}) with the derivative of $\delta^2(P_{h\perp})$
will contribute to the two spin observables.
For given ${\mathcal O}_{1,2}$ the integration over $\psi$ can be performed trivially: 
\begin{eqnarray} 
  \int d\psi L^{\mu\nu} {\mathcal O}_1 &=& 2\pi Q^2 \biggr ( -Y_M g^{\mu\nu} +Y_L P^\mu P^\nu  \biggr )  P_{h\perp}\cdot \tilde s_\perp, 
 \nonumber\\   
   \int d\psi L^{\mu\nu} {\mathcal O}_2 &=& 2\pi Q^4 Y_2 \biggr (P_{h\perp}^\mu \tilde s_\perp^\nu + P_{h\perp}^\nu \tilde s_\perp^\mu - \frac{2}{d-2}  g_\perp^{\mu\nu} P_{h\perp}\cdot \tilde s_\perp \biggr ).
\label{IPSI}      
\end{eqnarray}    
The above integrals should be understood as in $d$-dimensional space-time, 
i.e., $d\psi$ should be understood as $d\Omega_{d-2}$. Our final results are obtained by taking $d=4$. With $d=4$ 
we have for $Y_{L,M,2}$: 
\begin{eqnarray} 
  Y_M = \frac{1}{y^2} ( (1-y)^2 +1 ), \quad Y_L = \frac{4 x_B^2}{Q^2} ( (2-y)^2 + 2-2y), \quad Y_2 =\frac{(1-y)^2}{2 y^4}. 
\end{eqnarray}
With the tree-level result of the twist-3 part of $W^{\mu\nu}$  in Eq.(\ref{TreeW}) we have the results for the two observables:   
\begin{eqnarray}
\frac{ d\sigma \langle {\mathcal O}_1 \rangle }{d x_B d y d z_h} &=&  \pi \frac{ \alpha^2 y z_h }{ Q^2}  Y_M   
  \vert s_\perp \vert ^2  \int \frac{d x d z} { x z } \delta (1-\hat x) \delta (1-\hat z) \biggr [ \hat d(z) T_F(x,x) ( 1-\epsilon/2) -\epsilon h_1(x) \hat e_{\partial} (z) \biggr ], 
\nonumber\\  
   \frac{ d\sigma \langle {\mathcal O}_2 \rangle  }{d x_B d y d z_h} &=& \pi \alpha^2 y z_h  \vert s_\perp\vert^2 
   Y_2 \int \frac{d x d z} { x z} \delta (1-\hat x) \delta (1-\hat z)  
     2  h_1 (x) \hat e_\partial (z)   
     \biggr ( 3-\epsilon -\frac{2}{2-\epsilon}  \biggr ) , 
\label{TreeO}         
\end{eqnarray}
with 
\begin{equation} 
    \hat x =\frac{x_B}{x}, \quad  \hat z =\frac{z_h}{z}. 
\end{equation} 
From Eq.(\ref{TreeO}) it is observed that with $\epsilon=0$ the first observable only receives a chirality-even contribution, while the second one only receives a chirality-odd contribution. In Eq.(\ref{TreeO}) we have not included 
the contribution from charge-conjugated partonic process. This can be easily added to Eq.(\ref{TreeO}) through 
charge-conjugation.  The two observables have different dimensions in mass because the weights ${\mathcal O}_{1,2}$ 
are of different mass dimensions. This result in that different powers of $Q^2$ in the two observables.      
\par 
We will study one-loop corrections of these two observables. As discussed before, the virtual correction is determined 
by the quark form factor.  We will only need to calculate the real corrections.

\par\vskip20pt

\begin{figure}[hbt]
\begin{center}
\includegraphics[width=15cm]{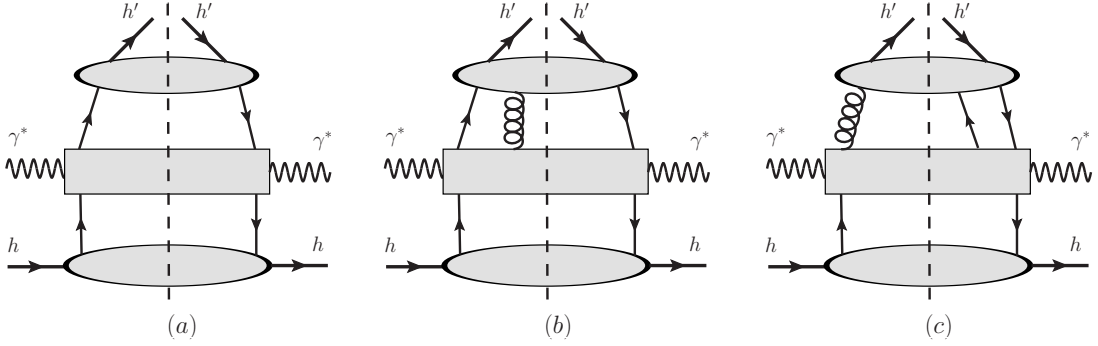}
\end{center}
\caption{Patterns of diagrams for  chiral-odd contributions to $W^{\mu\nu}$. The broken lines are the cuts.} 
\label{PCO}
\end{figure}

\noindent 
{\bf 3. One-Loop Chirality-Odd Correction}   
\par 
In this section we study one-loop chiral-odd corrections. They are from those diagrams which have the  general patterns given in Fig.1.  
In Fig.1. the bubbles in the lower part of diagrams represent the parton correlation given by the transversity distribution as in Eq.(\ref{LCH1}).  In Fig.1a the upper bubbles denote two parton correlations of fragmentation, while in Fig.1b and 1c they stand 
for corresponding three-parton correlations. The 
boxes in the middle of Fig.1 represent various parton scatterings. Contributions from complex conjugated diagrams of Fig.1b 
and Fig.1c  
should also be included.  In our case, we only need to consider the real corrections.  Hence, there is always one parton crossing 
the cut in the middle boxes. The virtual corrections are obtained as mentioned before.    

\par 
In Fig.1 we can in the first step make the projection for the lower bubbles with $h_1(x)$ as given in Eq.(\ref{LCH1}). 
The projection can be done in a frame in which the initial hadron moves in the $z$-direction, while the final hadron 
moves in an arbitrary direction. After the projection the contributions from Fig.1a and Fig.1b are:
\begin{eqnarray}
W^{\mu\nu}\biggr\vert_{1a} &=& \frac{1}{2N_c} \int \frac{ dx}{x} h_1(x) \int d^4 k_b {\rm Tr} \biggr \{ \gamma\cdot s_\perp 
  \gamma\cdot k_a 
      {\mathcal M}_{1a}^{\mu\nu}  (k_a,k_b) \Gamma_{1a} (k_b) \biggr \},       
\nonumber\\
W^{\mu\nu}\biggr\vert_{1b} &=& \frac{1}{2N_c} \int \frac{ dx}{x} h_1(x) \int d^4 k_b d^4 k_g {\rm Tr} \biggr \{ \gamma\cdot s_\perp 
  \gamma\cdot k_a 
      {\mathcal M}^{a,\mu\nu\alpha}_{1b}  (k_a,k_b, k_g) \Gamma_{1b}^{a,\beta}  (k_b,k_g) \biggr \} g_{\alpha\beta} ,
\end{eqnarray}
with $\Gamma_{1a,1b}$ for the upper bubbles given as:
\begin{eqnarray}
\Gamma_{1a,ij} (k_b) &=& \int \frac {d^4 \xi}{ (2\pi)^4} e^{-i\xi\cdot k_b}  \sum_X \langle  0\vert \psi_j (0)  \vert P_h, X \rangle  \langle  P_h, X \vert \bar \psi_i (\xi) \vert 0 \rangle,  
\nonumber\\
\Gamma_{1b,ij}^{a,\beta} (k_b,k_g) &=& \int \frac {d^4 \xi d^4 \xi_1 }{ (2\pi)^4 (2\pi)^4} e^{-i\xi\cdot k_b -ik_g\cdot\xi_1}  \sum_X \langle  0\vert \psi_j (0)  \vert P_h, X \rangle  \langle  P_h, X \vert \bar \psi_i (\xi)  G^{a,\beta} (\xi_1) \vert 0 \rangle,    
\end{eqnarray} 
where $ij$ stand for Dirac- and color indices. ${\mathcal M}_{1a}^{\mu\nu}$ and ${\mathcal M}_{1b}^{\mu\nu\alpha}$ stand for the boxes in the middle of Fig.1a and Fig.1b, respectively.  The contribution from Fig.1c takes a similar form. $k_a$ is the momentum of the quark from the lower bubbles and it is 
given by $k_a = x P$. It is noted that the projection, as given by the $(\cdots)$ in Eq.(\ref{LCH1}), can be written 
in a covariant form. After the projections from the lower bubbles we can do  projections from the upper bubbles 
in a frame in which the final hadron moves in the $-z$-direction. 

\par  
To find the contributions involving twist-3 parts of the upper bubbles one needs to perform collinear expansion for the parts represented 
by the middle boxes. The expansion includes the expansion of momenta carried by the parton lines connecting 
the boxes with the bubbles and projecting out perturbative parts from the middle boxes with the different parts 
of upper bubbles.  
It is noted that there are contributions at leading twist from Fig.1a. In general it is nontrivial to find 
the contributions at the next-to-leading twist if there are loops in the middle box. At twist-3 one of the two quark lines entering the upper bubble represents the bad component of the quark field. This component 
should be eliminated with equation of motion(See \cite{MaZhEE} and references therein).  
Since we deal here in fact only with tree-level diagrams
with a parton crossing the cut before its momentum is integrated over, the separation is rather easy. One can simply use the twist-3 part in the two-parton correlation in Eq.(\ref{2PFF}) for the upper bubble in Fig.1a. The diagrams represented by the middle box in Fig.1a
at the considered order of $\alpha_s$
are given in Fig.2. It is well-known that the existence of the transverse-spin part requires the existence of final state 
interactions in SIDIS, or nonzero absorptive part in scattering amplitudes. In the contributions of Fig.1a at the order of 
$\alpha_s$, the middle box can not have an absorptive part. The final state 
interactions can only appear in fragmentation.

\begin{figure}[hbt]
\begin{center}
\includegraphics[width=10cm]{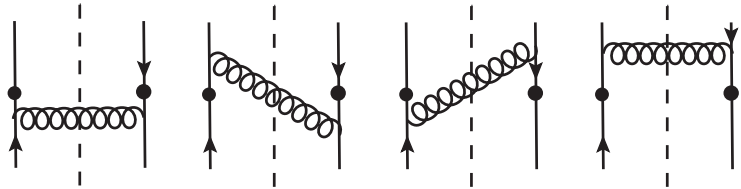}
\end{center}
\caption{Diagrams for  chiral-odd contributions,  represented by the middle box in Fig.1a. Black dots denote 
the insertion of the operator of electromagnetic current. } 
\label{2PCQ}
\end{figure}

\par 
The contributions from Fig.1b and Fig.1c are of twist-3 or higher twist. Because FF's from three-parton correlations have the support given
in Eq.(\ref{Z12R}) and are zero if any parton carries zero momentum fraction,  one finds at the order of $\alpha_s$ that the amplitudes of parton scatterings, represented by the left- or right parts of middle boxes, 
have no imaginary part, i.e., no physical cut. This is in contrast to chiral-even contributions studied in the next section. 
Therefore, the needed final state interactions only appear in FF's. The diagrams represented by the middle box 
in Fig.1b and Fig.1c are then given by Fig.3 and Fig.4, respectively.

\begin{figure}[hbt]
\begin{center}
\includegraphics[width=14cm]{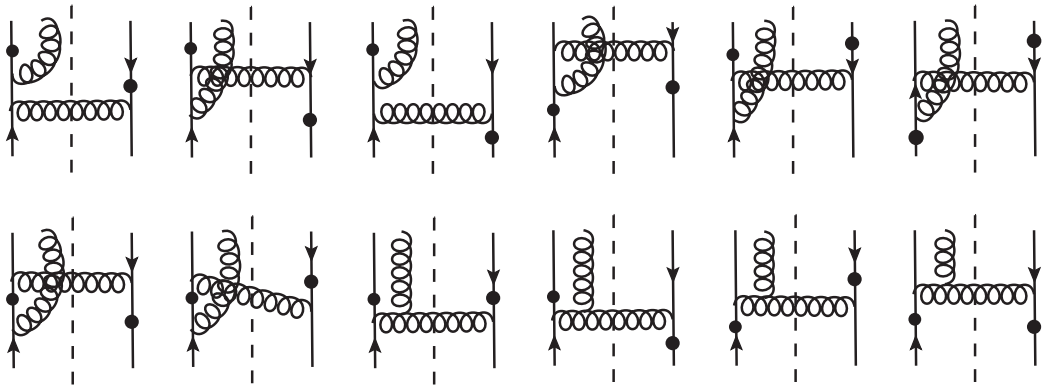}
\end{center}
\caption{Diagrams for  chiral-odd contributions, represented by the middle box in Fig.1b.   } 
\label{3PCQ}
\end{figure}

\begin{figure}[hbt]
\begin{center}
\includegraphics[width=14cm]{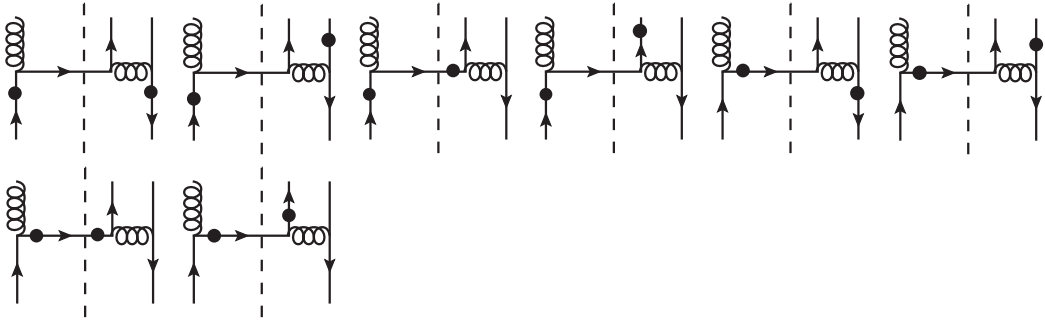}
\end{center}
\caption{Diagrams for  chiral-odd contributions, represented by the middle box in Fig.1c.  } 
\label{3PCG}
\end{figure}

\par 
The calculation is rather straightforward. One needs to perform the collinear expansion in the frame where the final hadron 
moves in the $-z$-direction. One essentially expands parton momenta $k_b$ and $k_g$ around the direction of $P_h^\mu$ 
and takes the large components of quark fields in the upper bubble of Fig.1b. Details about the expansion can be found in 
\cite{QiuSt, EKT}. 
 We first calculate $W^{\mu\nu}$ from Fig.1 with diagrams given explicitly in Fig.2, Fig.3 and Fig.4. The $U(1)$ gauge invariance is checked. From Fig.1a we obtain contributions 
involving $\hat e_\partial$ and $\hat e_I$. We use the relation in Eq.(\ref{RL2}) 
to express $\hat e_I$ with $\hat e_\partial$ 
and $\hat E_F$. With the obtained $W^{\mu\nu}$ we can calculate our spin-observables. Since there is only one parton in the intermediate state, the length of $P_{h\perp}$ is fixed as:
\begin{equation} 
   P_{h\perp} \cdot P_{h\perp} = - \frac{z z_h Q^2}{\hat x } (1-\hat x) (1 -\hat z).  
\end{equation} 
Therefore, the integration over $P_{h\perp}$ in Eq.(\ref{IOS}) can be performed easily.   
We have then the chiral-odd real corrections: 
\begin{eqnarray} 
\frac {d\sigma \langle {\mathcal O}_1\rangle}{ d x_B d y d z_h}\biggr\vert_{Re.}  &=& \frac{z_h\alpha_s \alpha^2 y}{4Q^2} \vert s_\perp\vert^2 F_D 
\int \frac{ d x d z}{x z} h_1 (x) \biggr \{ 
  \hat e_\partial (z) \biggr ( - 8 Y_M C_F \frac{2}{\epsilon}  \delta (1-\hat x) \delta (1-\hat z) 
   + {\mathcal A}_{1\sigma \partial}(\hat x,\hat z) \biggr ) 
\nonumber\\   
   &&  
 + 2 \int\frac{ d z_1}{z_1} \biggr [  {\rm Im} \hat E_F (z_1,z){\mathcal A}_{1\sigma F}(\hat x,\hat z,\hat z_1) 
  + {\rm Im } \hat E_G (z_1,z){\mathcal A}_{1\sigma G}(\hat x,\hat z,\hat z_1) \biggr] \biggr \},   
\nonumber\\
 \frac {d\sigma \langle {\mathcal O}_2\rangle}{ d x_B d y d z_h}\biggr\vert_{Re.} &=&\frac{z_h\alpha_s \alpha^2 y}{4} \vert s_\perp\vert^2 F_D 
 Y_2 \int \frac{ d x d z}{x z}  h_1 (x) \biggr \{ -\hat e_\partial (z)\frac{8}{\hat z}  C_F \biggr [ \biggr (  -2\biggr (\frac{2}{\epsilon} \biggr )^2 +3 \frac{2}{\epsilon} \biggr )\delta (1-\hat x) \delta (1-\hat z) 
\nonumber\\ 
     && + \delta (1-\hat x) \frac{2}{\epsilon} \frac{2 \hat z}{(1-\hat z)_+} +\delta (1-\hat z) \frac{2}{\epsilon}\frac{ 2 \hat x}{(1-\hat x)_+} \biggr ]   + \hat e_\partial (z) {\mathcal A}_{2\sigma \partial }(\hat x,\hat z)   + 2 \int \frac{ d z_1}{z_1} \biggr [ {\rm Im }  \hat E_F ( z_1,z)  
\nonumber\\     
    &&    \biggr ( \frac{2}{\epsilon} \delta (1-\hat x) \hat z \biggr ( C_F (\hat z -\hat z_1 -1) -\frac{N_c}{2} \frac{\hat z^2 +\hat z_1^2 -\hat z -\hat z_1}{ (\hat z-\hat z_1) (1-\hat z_1)} \biggr )\frac{4}{\hat z -\hat z_1}
    + {\mathcal A}_{2\sigma F}(\hat x,\hat z,\hat z_1) \biggr )
\nonumber\\
   && +  {\rm Im }\hat E_G ( z_1,z) \biggr ( \delta(1-\hat x) \frac{2}{\epsilon} \frac{ 4 C_F (\hat z-1)^2  }{ N_c  ( 1+\hat z_1 -\hat z)} + {\mathcal A}_{2\sigma G}(\hat x,\hat z,\hat z_1) \biggr )  \biggr ]  \biggr\}, 
\label{1LRCO}              
\end{eqnarray}     
where the poles of $1/\epsilon$ stand for collinear- or I.R. divergences. These divergences come from the momentum 
region where the unobserved parton in the intermediate state is soft or collinear to the initial- of final hadron. 
These divergences will be cancelled by those in the virtual parts or subtracted into parton distributions or FF's as we will show. The integrating ranges of $x$, $z$ and $z_1$  are given by:
\begin{equation} 
  \int d x = \int_{x_B}^1 dx, \quad \int d z = \int_{z_h}^1 d z, \quad \int d z_1 = \int_{z}^\infty d z_1. 
\end{equation}   
In Eq.(\ref{1LRCO}) we have already neglected those terms which are proportional to $\epsilon$ and will not 
contribute to our final results. $F_D$ and $\hat z_1$ are given by: 
\begin{equation}
    F_D =\left (\frac{4\pi\mu_c^2}{Q^2} \right )^{\epsilon/2} \frac{1}{\Gamma (1-\epsilon/2)}, \quad \hat z_1 =\frac {z_h}{z_1}  
\end{equation}
with $\mu_c$ as the scale associated with collinear divergences. 
In Eq.(\ref{1LRCO}) we have listed divergent contributions explicitly. The finite parts are given by functions 
${\mathcal A}'s$ which can be found in Appendix. 

\par 
As discussed before, the virtual correction to the derivative part of the hadronic tensor, hence to our observables, is given 
by the correction of the quark form factor. The correction is well-known. The virtual part can be simply obtained by multiplying our tree-level results in Eq.(\ref{TreeO}) with the factor: 
\begin{equation} 
      1 + \frac{\alpha_s C_F }{2\pi} F_D \biggr [ -2 \left (\frac{2}{\epsilon}\right )^2 -3 \left ( \frac{2}{\epsilon} \right ) -8 \biggr ] + {\mathcal O}(\alpha_s^2). 
\label{1LVF}      
\end{equation}       
Summing the divergent part in the real- and virtual part, we have the divergent parts of one-loop chirality-odd corrections: 
\begin{eqnarray} 
\frac {d\sigma \langle {\mathcal O}_1\rangle}{ d x_B d y d z_h}\biggr\vert_{Div.}  &=& 0, 
\nonumber\\
\frac {d\sigma \langle {\mathcal O}_2\rangle}{ d x_B d y d z_h}\biggr\vert_{Div.}  &=&
 z_h\alpha_s \alpha^2 y \vert s_\perp\vert^2 Y_2 F_D \left (\frac{2}{\epsilon} \right ) \int \frac{d x dz }{x z} h_1 (x) \biggr \{ 
    2 \hat e_\partial (z) C_F \biggr ( -3 \delta (1-\hat x) \delta (1-\hat z)  
\nonumber\\ 
   && - \frac{2\delta (1-\hat x)} {(1-\hat z)_+} -  \frac{ 2 \hat x\delta (1-\hat z)}{(1-\hat x)_+} \biggr )  
    + \delta (1-\hat x )\int \frac{ d z_1}{z_1} \biggr [  {\rm Im } \hat E_F ( z_1,z)  \frac{ 2 \hat z}{\hat z- \hat z_1}
\nonumber\\
   &&  \biggr ( 
      C_F ( \hat z -\hat z_1 -1) -\frac{N_c}{2} \frac{\hat z^2 +\hat z_1^2 -\hat z -\hat z_1}{ (\hat z -\hat z_1)(1-\hat z_1)} 
     \biggr )
 + {\rm Im} \hat E_G( z_1,z)  \frac{  2 C_F (\hat z -1)^2}{ N_c (1+\hat z_1 -\hat z)} \biggr ] \biggr \}.
\label{1LCOD}   
\end{eqnarray}
In the sums, the infrared divergences are cancelled. The remaining divergences are collinear ones. The one-loop correction to our first spin observables is finite. The correction to the second observable  
contains collinear divergences. 

\par 
It should be noted that the contributions from exchanging collinear partons are in fact already included in 
the parton distributions and FF's of the tree-level results in Eq.(\ref{TreeO}), or they are already contained in the lower- and upper bubbles in Fig.5.  In order to avoid double counting, we should subtract these contributions from the one-loop corrections calculated in the above. The subtraction at one-loop level 
can be easily done with the replacement in the tree-level results in Eq.(\ref{TreeO}):
\begin{equation} 
  h_1 (x) \to h_1 (x)-\Delta h_1(x), \quad \hat e_\partial (x) \to \hat e_\partial(x) - \Delta \hat e_\partial (x)
\end{equation} 
and the contributions which need to be subtracted are: 
\begin{eqnarray} 
 \Delta \frac {d\sigma \langle {\mathcal O}_1\rangle}{ d x_B d y d z_h}  &=& \frac{\pi z_h y \alpha^2}{Q^2} Y_M \vert s_\perp\vert^2  \epsilon  \biggr ( \Delta h_1 (x_B) \hat e_\partial (z_h) + h_1 (x_B) \Delta \hat e_\partial (z_h) \biggr ) , 
\nonumber\\
 \Delta \frac {d\sigma \langle {\mathcal O}_2\rangle}{ d x_B d y d z_h}  &=& -\pi z_h y \alpha^2 Y_2 \vert s_\perp\vert^2 
   \biggr ( 4 - 3\epsilon  \biggr )  \biggr ( \Delta h_1 (x_B) \hat e_\partial (z_h) + h_1 (x_B) \Delta \hat e_\partial (z_h) \biggr ). 
\label{SUBCO}    
\end{eqnarray}                      
These contributions should be added in our final results to avoid the double counting. 
\par 
In the case with dimensional regularization for collinear divergences of massless partons, the quantities $\Delta h_1$ and 
$\Delta e_\partial$ are determined by the evolution of $h_1$ and $\hat e_\partial$, respectively.  The evolution of twist-3 FF's have been studied in \cite{MaZhEE,BeKu,KangEE}. The evolution of $h_1$ 
can be found in \cite{WVTE}. The evolution of $\hat e_\partial$ can be found in \cite{MaZhEE,KangEE}. According to these results 
 $\Delta h_1$ and 
$\Delta e_\partial$ are given by:
\begin{eqnarray} 
\Delta h_1(x_B) &=& \frac{\alpha_s}{2\pi} \biggr (-\frac{2}{\epsilon} + \ln \frac{e^\gamma \mu^2}{4\pi \mu_c^2} \biggr )  
   \int \frac{dx}{x}  C_F \biggr ( \frac {2\hat x}{(1-\hat x )_+ } +\frac{3}{2} \delta (1-\hat x) \biggr ) h_1 (x)
\nonumber\\ 
    &=& \frac{\alpha_s}{2\pi} \biggr (-\frac{2}{\epsilon} + \ln \frac{e^\gamma \mu^2}{4\pi \mu_c^2} \biggr ) 
       \biggr ( P_{\perp q}\otimes h_1 \biggr ) (x_B),      
\nonumber\\
\Delta \hat e_\partial (z_h) &=& \frac{\alpha_s}{2\pi} \biggr (-\frac{2}{\epsilon} + \ln \frac{e^\gamma \mu^2}{4\pi \mu_c^2} \biggr ) \int \frac{ dz}{z} \biggr \{ \hat e_\partial (z) C_F \biggr ( \frac{2}{(1-\hat z)_+} +\frac{3}{2} \delta (1-\hat z) \biggr ) 
  - \int \frac{ d z_1}{z_1} \biggr [  {\rm Im } \hat E_F ( z_1,z) 
\nonumber\\
   &&  \frac{  \hat z}{\hat z- \hat z_1}\biggr ( 
      C_F ( \hat z -\hat z_1 -1) -\frac{N_c}{2} \frac{\hat z^2 +\hat z_1^2 -\hat z -\hat z_1}{ (\hat z -\hat z_1)(1-\hat z_1)} 
     \biggr )
 + {\rm Im} \hat E_G( z_1,z)  \frac{  C_F (\hat z -1)^2}{ N_c (1+\hat z_1 -\hat z)} \biggr ] \biggr \}
\nonumber\\
   &=& \frac{\alpha_s}{2\pi} \biggr (-\frac{2}{\epsilon} + \ln \frac{e^\gamma \mu^2}{4\pi \mu_c^2} \biggr ) 
       \biggr ( {\mathcal F}_\partial \otimes \hat e_\partial + {\mathcal F}_F \otimes \hat E_F + {\mathcal F}_G \otimes 
        \hat E_G \biggr ) (z_h),  
\label{Dh1ep}  
\end{eqnarray} 
where the poles in $\epsilon$ stand for collinear divergence and $\mu$ is the renormalization scale. In Eq.(\ref{Dh1ep}) 
we have introduced four evolution kernels $P_{\perp q}$, $ {\mathcal F}_\partial$ , ${\mathcal F}_F$ and ${\mathcal F}_G$ 
for a short notation. 
Taking the derivative 
of $\Delta h_1$ and 
$\Delta e_\partial$ with respect to $\ln \mu$ one obtains the evolution of $h_1$ and $\hat e_\partial$, respectively.

\par 
Substituting the results in Eq.(\ref{Dh1ep}) into the contributions in Eq.(\ref{SUBCO}) and adding them to the 
divergent part of the one-loop corrections in Eq.(\ref{1LCOD}), one can realize that the divergences represented 
by the poles in $\epsilon$ are cancelled. The final results of chirality-odd contributions are finite.  We will present 
the final results in Sect. 5.

\begin{figure}[hbt]
\begin{center}
\includegraphics[width=15cm]{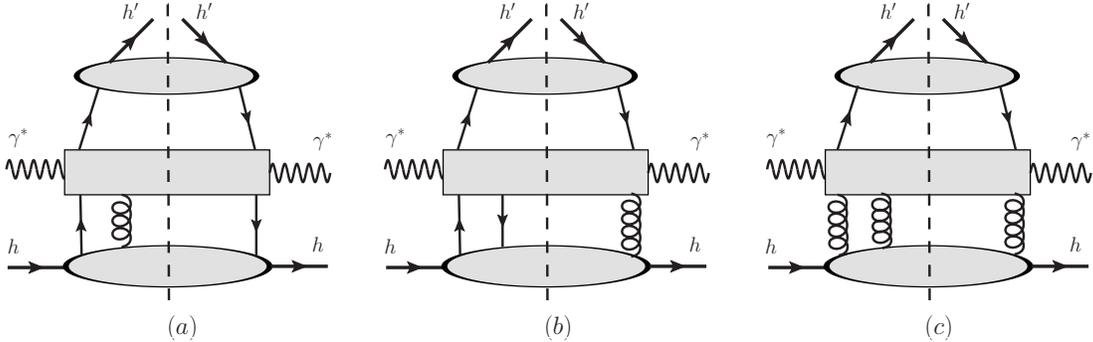}
\end{center}
\caption{Patterns of diagrams for  chiral-even contributions to $W^{\mu\nu}$. The broken lines are the cuts.} 
\label{PCO}
\end{figure}
     
\par\vskip20pt
\noindent 
{\bf 4. One-Loop Chirality-Even Correction}   
\par 
In this section we study the one-loop chirality-even corrections. As discussed before, we only need to calculate 
the real correction. The virtual correction  is given by the one-loop correction of the quark form factor. 
The chirality-even contributions involve twist-2 parton FF's and twist-3 parton distributions of the 
initial hadron. 
\par 
In the case of quark fragmentation, the contributions are from these diagrams, whose general structure can be represented 
by Fig.5.  To calculate, e.g., the contribution from Fig.5a, one can in the first step project out the contribution of 
the twist-2 quark FF from the upper bubble. After the projection the contribution to $W^{\mu\nu}$ can be written as:
\begin{eqnarray}
W^{\mu\nu}\biggr\vert_{5a} =  \int \frac{ dz}{z^{d-3}} \hat d(z) \int d^d k_a d^d k_g  {\rm Tr} \biggr \{ \gamma\cdot P_h   
      {\mathcal M}_{5a}^{a,\mu\nu\alpha}   (k_a, k_g, k_b) \Gamma_{5a}^{a,\beta} (k_a,k_g) \biggr \} g_{\alpha\beta} ,
\end{eqnarray}              
with $\Gamma_{5a}$ for the lower bubble in Fig.5a given as:
\begin{eqnarray}
\Gamma_{5a,ij}^{a,\beta} (k_a,k_g ) = \int \frac {d^d \xi_a d^d \xi_g }{ (2\pi)^{2d}} e^{i\xi_a\cdot k_a + i \xi_g \cdot k_g }   \langle  P, s_\perp \vert \bar  \psi_j (0) G^{a,\beta}(\xi_g) \psi_i (\xi_a ) \vert P, s_\perp  \rangle
\end{eqnarray} 
and ${\mathcal M}_{5a}^{a,\mu}$ standing for the middle box in Fig.5a. In the above contribution the collinear expansion 
relevant to the produced hadron is performed and the momentum carried by the quark lines between the middle box and the 
upper bubble is given by $k_b = P_h/z$. $k_a$ and $k_g$ are the momenta carried by the quark- and gluon lines in the low-left 
part of Fig.5a, respectively.       
\par 
At the order we consider, there is always one parton contained in the middle boxes crossing the cut. Unlike the case of chiral-odd contributions studied before, where the final-state interactions only appear 
in parton fragmentation, the final-state interactions in chirality-even contributions can only appear in the middle boxes, i.e., in the hard scatterings. This is due to that twist-2 parton FF's do not contain final-state interactions.  
The amplitudes represented by the left parts of the middle boxes 
have nonzero absorptive part, or the left parts contain a physical cut implicitly. Because of the cut, one of the parton lines 
in the low-left part of diagrams in Fig.5 can carry zero momentum. The resulted contributions are called as soft-quark- or 
soft-gluon pole contributions. There are contributions in which none  of the parton lines carry zero momentum. These contributions are called as hard-pole contributions. There is also the case that the final hadron is produced through 
gluon fragmentation. In this case, there are types of diagrams similar to Fig.5.        

\begin{figure}[hbt]
\begin{center}
\includegraphics[width=11cm]{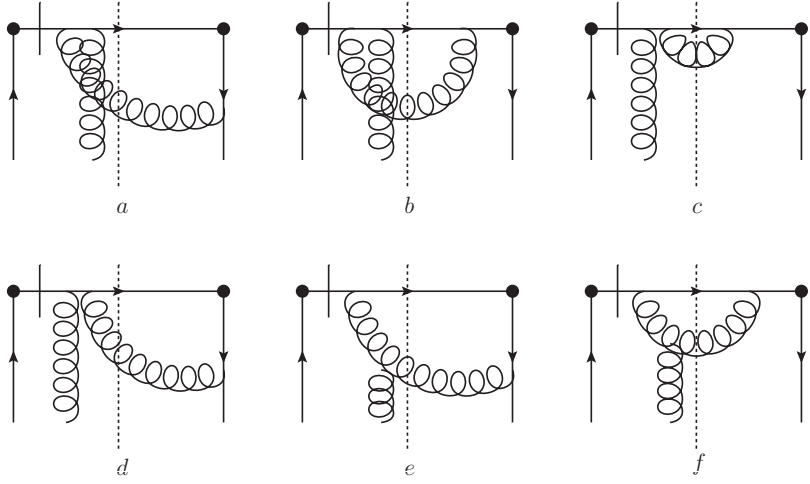}
\end{center}
\caption{Feynman diagrams for the hard-pole contribution with quark- or gluon fragmentation. } 
\label{HPQG}
\end{figure}
\par

\par 
It has been studied in detail how to make the collinear expansion in a gauge-invariant way in \cite{EKT2} and 
how to calculate the soft-gluon-pole contributions with the master formulas found in \cite{KoTa,KoTa2,KTY}. 
Employing these technics various contributions can be calculated in a straightforward way.   
Therefore, we will not give the detail of how these calculations are done.  We first discuss the contributions of hard-poles. For the case of quark fragmentation, the hard scattering part represented
by the middle box in Fig.5a is given by the diagrams in Fig.6, where the quark propagators with a short bar implies the cut 
for the absorptive part, i.e., only the absorptive part of the propagators is taken into account. The dispersive part 
will not contribute. Fig.6 stands for two cases, the final hadron can be produced from quark- or gluon fragmentation.   
The contributions from Fig.6 are:  
\begin{eqnarray}
\frac{ d \sigma \langle {\mathcal O}_1\rangle}{ d x_B dy d z_h} \biggr\vert_{Fig.6} &=& \frac{z_h y\alpha^2 \alpha_s}{4 Q^2} 
  \vert s_\perp\vert^2 F_D \int \frac{d x d z}{xz}\biggr \{  \hat d (z) 
    \biggr [  Y_M T_F (x,x_B) \biggr (  2 N_c\left ( \frac{2}{\epsilon}\right ) ^2 \delta (1-\hat x) \delta (1-\hat z) 
\nonumber\\
   && -\frac{ 1 +\hat z (N_c^2-1)}{\hat z N_c} \frac{2}{\epsilon}  \delta (1-\hat x) \frac{1+\hat z^2}{ (1-\hat z)_+}     -N_c \delta (1-\hat z) \frac{2}{\epsilon}\frac{1+\hat x}{(1-\hat x)_+} \biggr ) 
\nonumber\\     
     &&   + Y_M T_{\Delta} ( x, x_B) N_c \delta (1-\hat z) \frac{2}{\epsilon} 
+ T_F (x,x_B) {\mathcal A}_{1hq} (\hat x, \hat z) 
      + T_\Delta  (x,x_B) {\mathcal B}_{1hq} (\hat x, \hat z)  \biggr ] 
\nonumber\\      
     && + \hat g (z)   \biggr [  
      Y_M T_F (x,x_B)
   \frac{\hat z + N_c^2 (1-\hat z)}{ \hat z^2 N_c} \delta (1-\hat x) \frac{2}{\epsilon} (2 -2\hat z + z^2) 
\nonumber\\     
     && + T_F (x,x_B) {\mathcal A}_{1hg} (\hat x, \hat z)  
      + T_\Delta  (x,x_B) {\mathcal B}_{1hg} (\hat x, \hat z)  \biggr ]  \biggr \},          
\nonumber\\
\frac{ d \sigma \langle {\mathcal O}_2\rangle}{ d x_B dy d z_h} \biggr\vert_{Fig.6} &=& 0. 
\label{HP1} 
\end{eqnarray} 
Here we only list the divergent results explicitly. ${\mathcal A}_{1hi} (\hat x, \hat z)$ and ${\mathcal B}_{1hi} (\hat x, \hat z)$ for $i=q,g$ are finite functions given in Appendix. In the following, we will give various contributions in the same way as Eq.(\ref{HP1}). 
From Fig.6 our second observable receives no contributions.

\begin{figure}[hbt]
\begin{center}
\includegraphics[width=13cm]{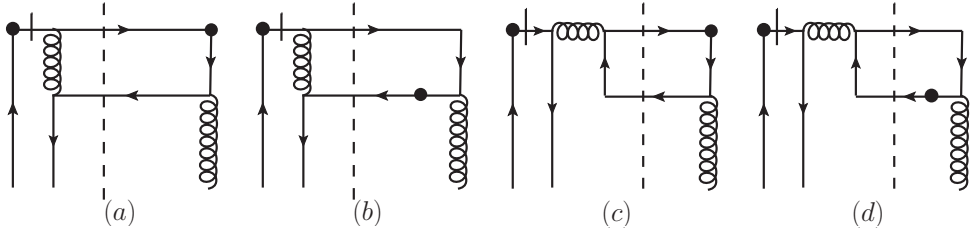}
\end{center}
\caption{Feynman diagrams for the hard scattering represented by the middle part of Fig.5b. Their contributions are of hard-pole with quark- or antiquark fragmentation.  } 
\label{HPQQB}
\end{figure}
\par 
There are partonic processes as the forward scattering in which a $q\bar q$ pair participates as indicated by Fig.5b.  Their contributions are given by Fig.7. These contributions are of hard-pole. The contributions from Fig.7 are:
\begin{eqnarray}
\frac{ d \sigma \langle {\mathcal O}_1\rangle}{ d x_B dy d z_h} \biggr\vert_{Fig.7} &=& \frac{z_h y\alpha^2 \alpha_s}{4 Q^2} 
  \vert s_\perp\vert^2 F_D \int \frac{d x d z}{xz}  \biggr \{  \hat d (z) \biggr [ 
       -\frac{1}{N_c} Y_M T_F (x_B-x,x_B) \frac{2}{\epsilon} \delta (1-\hat z)  (1-2\hat x) 
\nonumber\\     
   &&  + \frac{1}{N_c} Y_M T_{\Delta} (x_B-x,x_B)  \frac{2}{\epsilon} \delta (1-\hat z) \biggr ] 
   +T_F (x_B-x,x_B) \biggr ( \hat d(z)  {\mathcal C}_{1Fq} (\hat x,\hat z) 
\nonumber\\   
   &&  -\hat d(-z) {\mathcal C}_{1F\bar q} (\hat x,\hat z)  \biggr )   
  + T_{\Delta} (x_B-x,x_B)  \biggr ( \hat d(z) {\mathcal C}_{1Dq} (\hat x, \hat z) - \hat d(- z){\mathcal C}_{1D\bar q} (\hat x, \hat z) \biggr )             \biggr\},
\nonumber\\
\frac{ d \sigma \langle {\mathcal O}_2\rangle}{ d x_B dy d z_h} \biggr\vert_{Fig.7} &=& \frac{z_h y\alpha^2 \alpha_s}{4} 
  \vert s_\perp\vert^2 Y_2 \int \frac{d x d z}{xz}  \biggr \{     T_F (x_B-x,x_B) \biggr ( \hat d(z)  {\mathcal C}_{2Fq} (\hat x,\hat z) 
\nonumber\\   
   && -\hat d(-z) {\mathcal C}_{2F\bar q} (\hat x,\hat z)  \biggr )   
  + T_{\Delta} (x_B-x,x_B)  \biggr ( \hat d(z) {\mathcal C}_{2Dq} (\hat x, \hat z) -\hat d(-z){\mathcal C}_{1D\bar q} (\hat x, \hat z) \biggr )             \biggr\}.        
\label{HP2} 
\end{eqnarray}  
In the above contributions, one of variables of twist-3 parton distributions $T_{F,\Delta}(x_1,x_2)$ is negative, its absolute value is the 
momentum fraction of the anti-quark. The anti-quark FF is given by $-\hat d(-z)$.  
The contributions from Fig.7 to our second observable is nonzero but finite.

\begin{figure}[hbt]
\begin{center}
\includegraphics[width=14cm]{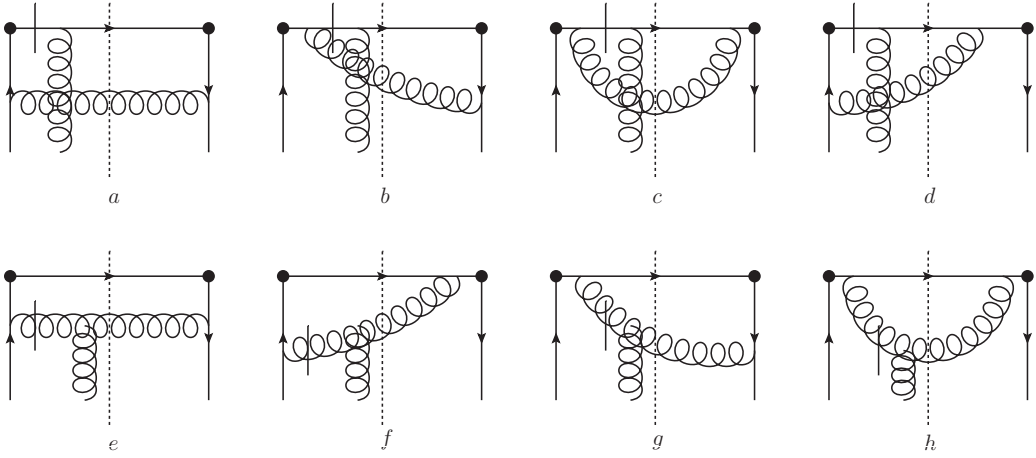}
\end{center}
\caption{Feynman diagrams for the hard scattering represented by the middle part of Fig.5a. Their contributions are of soft-gluon-pole. Diagrams in the first(second) row are of quark(gluon) fragmentation.  } 
\label{HPQQB}
\end{figure}

\par 
There are soft-pole 
contributions, in which one of initial partons carries zero momentum. This parton can be a gluon or quark. The soft-gluon pole contributions can be calculated conveniently with the master formulas found in \cite{KoTa,KoTa2,KTY}. From the type of diagrams of Fig.5a there are soft-gluon-pole contributions. They are given 
by diagrams in Fig.8. In Fig.8 the diagrams of the first row are for quark fragmentation, those in the second row 
are for gluon fragmentation. The total contributions from Fig.8 are: 
\begin{eqnarray}
\frac{ d \sigma \langle {\mathcal O}_1\rangle}{ d x_B dy d z_h} \biggr\vert_{Fig.8} &=& \frac{z_h y\alpha^2 \alpha_s}{4 Q^2} 
  \vert s_\perp\vert^2 F_D \int \frac{d x d z}{xz}  T_F (x,x)  \biggr \{ \hat d(z) \biggr [ 
     \frac{1}{\hat z N_c} Y_M  \biggr (  (\epsilon -2 ) \left (\frac{2}{\epsilon}\right )^2 
      \delta (1-\hat x) \delta (1-\hat z) 
\nonumber\\
     && + \frac{2}{\epsilon} \frac{1+\hat z^2}{(1-\hat z)_+} \delta (1-\hat x) 
     + \frac{2}{\epsilon} \frac{1+\hat x^2}{(1-\hat x)_+} \delta (1-\hat z) \biggr ) 
     + {\mathcal D}_{1q}(\hat x,\hat z) \biggr ] 
\nonumber\\        
      &&  + \hat g (z)   \biggr [ 
     \frac{N_c}{\hat z^2 } Y_M  \biggr ( - \frac{2}{\epsilon} \delta (1-\hat x) ( 2 -2 \hat z +\hat z^2)  \biggr ) 
+  {\mathcal D}_{1g}(\hat x,\hat z) 
         \biggr ]           
      \biggr \},    
\nonumber\\
\frac{ d \sigma \langle {\mathcal O}_2\rangle}{ d x_B dy d z_h} \biggr\vert_{Fig.8} &=& 0. 
\label{HSG} 
\end{eqnarray}    
Our second observable does not receive contributions from Fig.8. 

\begin{figure}[hbt]
\begin{center}
\includegraphics[width=13cm]{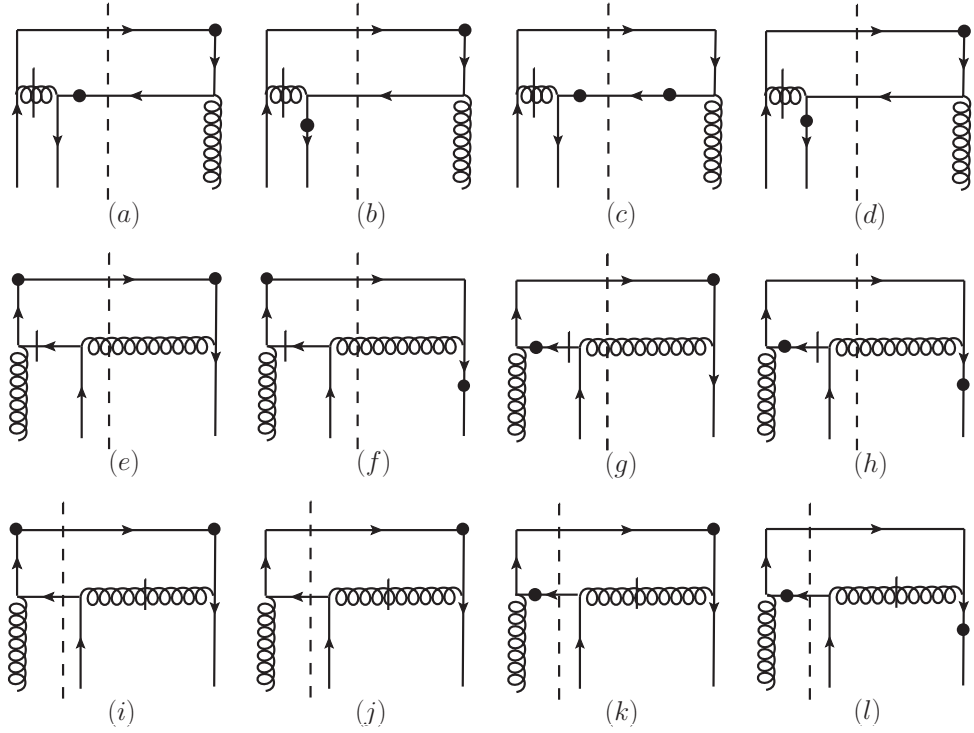}
\end{center}
\caption{Feynman diagrams for the hard scattering represented by the middle part of Fig.5b. Their contributions are of soft-quark-pole.  } 
\label{HPQQB}
\end{figure}

\par 
From the type of diagrams of Fig.5b there are soft-quark-pole contributions. The diagrams are given in Fig.9. 
There are contributions involving quark-, antiquark and gluon fragmentation functions. The contributions from gluon 
fragmentation are represented by the diagrams in the second row. Those from antiquark fragmentation are given by the first- 
and third row. Contributions with quark fragmentation are from all three rows of Fig.9. But the contributions from the second 
row cancel those from the complex conjugated diagrams of third row. This is because the diagrams in the third row are from the second row 
by cutting the unobserved parton lines in different ways. It is easy to show the cancellation.  Therefore, the contributions of quark fragmentation are only 
from the first row. We have:     
\begin{eqnarray}
\frac{ d \sigma \langle {\mathcal O}_1\rangle}{ d x_B dy d z_h} \biggr\vert_{Fig.9} &=& \frac{z_h y\alpha^2 \alpha_s}{4 Q^2} 
  \vert s_\perp\vert^2  \int \frac{d x d z}{xz}  \biggr \{ \hat d (z)  \biggr [ T_F(-x,0) {\mathcal E}_{1Fq} (\hat x,\hat z) 
 +T_\Delta (-x,0) {\mathcal E}_{1\Delta q} (\hat x,\hat z)  \biggr ]
\nonumber\\
     && -\hat d (-z)  \biggr [ T_F(0,x) {\mathcal E}_{1F\bar q} (\hat x,\hat z) 
 +T_\Delta (0,x) {\mathcal E}_{1\Delta\bar  q} (\hat x,\hat z)  \biggr ] 
\nonumber\\
   && + \hat g(z) \biggr [ T_F(x,0) {\mathcal E}_{1Fg} (\hat x,\hat z)   
 +T_\Delta (x,0) {\mathcal E}_{1\Delta g} (\hat x,\hat z)  \biggr ] \biggr \}, 
\nonumber\\
\frac{ d \sigma \langle {\mathcal O}_2\rangle}{ d x_B dy d z_h} \biggr\vert_{Fig.9} &=& \frac{z_h y\alpha^2 \alpha_s}{4 } 
  \vert s_\perp\vert^2 Y_2  \int \frac{d x d z}{xz}  \biggr \{ \hat d (z)  \biggr [ T_F(-x,0) {\mathcal E}_{2Fq} (\hat x,\hat z)  + T_\Delta (-x,0) {\mathcal E}_{2\Delta q} (\hat x,\hat z) \biggr ]
\nonumber\\
     && -\hat d (-z)  \biggr [ T_F(0,x) {\mathcal E}_{2F\bar q} (\hat x,\hat z)  
      +T_\Delta (0,x) {\mathcal E}_{2\Delta\bar  q} (\hat x,\hat z)  \biggr ]  
\nonumber\\
   && + \hat g(z) \biggr [ T_F(x,0) {\mathcal E}_{2Fg} (\hat x,\hat z) +  T_\Delta (x,0) {\mathcal E}_{2\Delta g} (\hat x,\hat z)  \biggr ] \biggr \}.  
\label{HSQ} 
\end{eqnarray}    
The soft-quark-pole contributions are finite.

\par

\begin{figure}[hbt]
\begin{center}
\includegraphics[width=14cm]{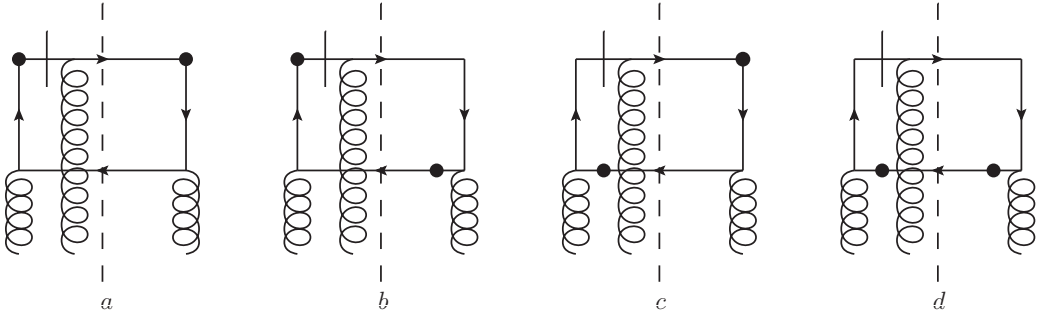}
\end{center}
\caption{Feynman diagrams for the hard scattering represented by the middle part of Fig.5c. Their contributions are of soft-gluon-pole   } 
\label{HP3G}
\end{figure}

At the order we consider, there are contributions involving gluonic twist-3 parton distributions defined 
in Eq.(\ref{3GT3}). These contributions are represented by the type of diagrams specified in Fig.5c. Since there are three gluons 
exchanged between the middle box and the lower bubble, the Bose-symmetry should be kept.  
For this it is convenient to use the symmetric notation in \cite{KTY,Ji3G,BKTY} for the twist-3 gluonic matrix elements in the 
calculation and to express the final result with the gluonic distributions in Eq.(\ref{3GT3}). 
In the symmetric notation the matrix element of twist-3 gluonic operator can be parameterized as:
\begin{eqnarray} 
&& \frac{i^3 g_s}{P^+}  \int \frac{d\lambda_1}{2\pi }\frac{d\lambda_2}{2\pi} 
   e^{i\lambda_1 x_1 P^+ + i\lambda_2 (x_2-x_1)P^+} \langle P,s_\perp  \vert G^{a,+\alpha} 
   (\lambda_1 n)  G^{c,+\gamma}(\lambda_2 n) G^{b,+\beta} (0) \vert P,s_\perp  \rangle 
\nonumber\\
   && = \frac{N_c}{(N_c^2-1)(N_c^2-4)} d^{abc} O^{\alpha\beta\gamma}(x_1,x_2) - \frac{i }{N_c(N_c^2-1) } f^{abc} N^{\alpha\beta\gamma}(x_1,x_2), 
\label{NO3G}    
\end{eqnarray} 
where all indices $\alpha,\beta$ and $\gamma$ are transverse. From Bose-symmetry and covariance                
the two tensors take the form: 
\begin{eqnarray}
O^{\alpha\beta\gamma}(x_1,x_2) &=& -2 i \biggr [ O(x_1,x_2) g^{\alpha\beta} \tilde s_\perp^\gamma + 
  O(x_2,x_2-x_1) g^{\beta\gamma} \tilde s_\perp^\alpha + O(x_1,x_1-x_2) g^{\gamma\alpha} \tilde s_\perp^\beta \biggr ], 
\nonumber\\
N^{\alpha\beta\gamma}(x_1,x_2) &=& -2 i \biggr [ N(x_1,x_2) g^{\alpha\beta} \tilde s_\perp^\gamma - 
  N(x_2,x_2-x_1) g^{\beta\gamma} \tilde s_\perp^\alpha - N(x_1,x_1-x_2) g^{\gamma\alpha} \tilde s_\perp^\beta \biggr ],
\label{NOG3}           
\end{eqnarray}
with the properties of the function $O$ and $N$ 
\begin{eqnarray} 
    && O(x_1,x_2) =O(x_2,x_1),\quad O(x_1,x_2) = O(-x_1,-x_2), 
\quad  N(x_1,x_2) = N(x_2,x_1),
\nonumber\\ 
   && N(x_1,x_2) = -N(-x_1,-x_2). 
\end{eqnarray}      
The function $O$ and $N$  are related to those defined in Eq.(\ref{3GT3}) as 
\begin{eqnarray} 
 T_G^{(f)}(x_1,x_2) &=& 2\pi \biggr ( (d-2) N(x_1,x_2) - N(x_2,x_2-x_1) - N(x_1,x_1-x_2) \biggr ), 
\nonumber\\  
   T_G^{(d)}(x_1,x_2) &=& 2\pi \biggr ( (d-2)  O (x_1,x_2) + O(x_2,x_2-x_1) +O (x_1,x_1-x_2) \biggr ).  
\label{NOTG}             
\end{eqnarray}
It should be noted that the relations depend on $d=4-\epsilon$. It will affect on the subtraction discussed later. 
\par 
With the symmetric notation, the middle box of Fig.5c is given by diagrams in Fig.10. The contributions from Fig.5c 
are of soft-gluon-pole, i.e., one of the three gluon-lines carries zero momentum. One can use the so-called master formula 
in \cite{KoTa,KoTa2,KTY} to calculate the twist-3 contributions. From Fig.10 we have: 
\begin{eqnarray}
\frac{ d \sigma \langle {\mathcal O}_1\rangle}{ d x_B dy d z_h} \biggr\vert_{Fig.10} &=& \frac{z_h y\alpha^2 \alpha_s}{4 Q^2} 
  \vert s_\perp\vert^2 F_D (-4\pi )  \int \frac{d x d z}{x^2 z}  \hat d(z)   \biggr \{  \biggr ( 
    O(x,x) + N(x,x) + O(x,0)- N(x,0) \biggr )  
\nonumber\\     
   &&  \biggr [ Y_M \frac{2}{\epsilon} \delta (1-\hat z) (1-2\hat x + 2 \hat x^2) + Y_L {\mathcal F}_{1L}(\hat x, \hat z)\biggr ] + Y_M \biggr (O(x,x)+N(x,x)  \biggr ) 
\nonumber\\
   &&   {\mathcal F}_{1M+} (\hat x,\hat z) + Y_M \biggr (O(x,0) - N(x,0) \biggr ) {\mathcal F}_{1M-} (\hat x,\hat z)
       \biggr \},       
\nonumber\\
\frac{ d \sigma \langle {\mathcal O}_2\rangle}{ d x_B dy d z_h} \biggr\vert_{Fig.10} &=& \frac{z_h y\alpha^2 \alpha_s}{4 } 
  \vert s_\perp\vert^2  Y_2 (-4\pi )  \int \frac{d x d z}{x^2 z}  \hat d(z)   \biggr  ( O(x,0)- N(x,0) \biggr ) \frac{4 \hat x^2}{\hat z}. 
\label{H3G} 
\end{eqnarray}    
We observe that the contribution from Fig.10 to our second observable is finite. 

\par 
The real part of one-loop chirality-even correction to our observables is the sum 
of the contributions listed in Eq.(\ref{HP1}, \ref{HP2}, \ref{HSG}, \ref{HSQ}, \ref{H3G}). The virtual part, 
as mentioned before, is obtained by multiplying 
our tree-level results in Eq.(\ref{TreeO}) by the factor in Eq.(\ref{1LVF}). Since our second observable at tree-level does not contain 
chirality-even part, its virtual part of the chirality-even part is zero. Adding all divergent contributions 
together, we obtain the sum:
\begin{eqnarray}
\frac{ d \sigma \langle {\mathcal O}_1\rangle}{ d x_B dy d z_h} \biggr\vert_{Div.} &=& \frac{z_h y\alpha^2 \alpha_s}{2 Q^2} 
  \vert s_\perp\vert^2 Y_M F_D   \int \frac{d x d z}{x z} \left (-\frac{2}{\epsilon} \right ) 
  \biggr \{ \delta (1-\hat x) T_F (x,x) \biggr ( {\mathcal P}_{qq} (\hat z) \hat d(z) 
\nonumber\\  
  && + {\mathcal P}_{gq} (\hat z) \hat g(z) \biggr ) 
     +\delta (1-\hat z) \hat d(z) \biggr [ \biggr (  {\mathcal P}_{qq}(\hat x) -\frac{N_c}{2} \frac{1+\hat x^2}{(1-\hat x)_+}
      - N_c \delta (1-\hat x) \biggr )  T_F (x,x) 
\nonumber\\     
     && +\frac{N_c}{2} \frac{ 1 +\hat x}{(1-\hat x)_+} T_F (x,x_B)  +\frac{N_c}{2} T_\Delta (x_B,x) +\frac{1}{2 N_c}  T_\Delta (x_B,x_B-x)  
\nonumber\\     
     && +\frac{1}{2 N_c} (1-2\hat x) T_F(x_B,x_B-x)  
     -\frac{1}{2 x} ( 1-2\hat x +2\hat x^2) T_{G+}(x,x) \biggr ] \biggr \},       
\nonumber\\
\frac{ d \sigma \langle {\mathcal O}_2\rangle}{ d x_B dy d z_h} \biggr\vert_{Div. } &=& 0, 
\label{DIV} 
\end{eqnarray}    
with the standard splitting functions:
\begin{eqnarray}
  {\mathcal P}_{qq}(z) = C_F \biggr ( \frac {1+z^2}{(1-z)_+} +\frac{3}{2}\delta (1-z) \biggr ), \quad 
  {\mathcal P}_{gq}(z) = C_F  \frac {2-2z+z^2}{z} . 
\end{eqnarray}
In Eq.(\ref{DIV}) and in our final results the contributions from gluonic twist-3 distributions can be conveniently 
expressed by two combinations $T_{G\pm}(x_1,x_2)$ of gluonic twist-3 distributions. They are defined as:
\begin{eqnarray} 
    T_{G\pm} (x_1,x_2) = T_G^{(f)}(x_1,x_2) \pm T_G^{(d)} (x_1,x_2). 
\end{eqnarray}
In Eq.(\ref{DIV}) the divergent gluonic contribution is only related to $T_{G+}(x,x)$, which are given by
those twist-3 gluonic distributions in Eq.(\ref{NOG3}) as: 
\begin{eqnarray} 
   T_{G+} (x,x) &=& 2\pi \biggr [ 2 \biggr ( O(x,x)+N(x,x)+O(x,0) - N(x,0) \biggr ) - \epsilon \biggr ( O(x,x) 
           +N(x,x) \biggr ) \biggr ].             
\end{eqnarray} 
Inspecting Eq.(\ref{DIV}), we find that our second observable is finite, while the first one is divergent. The divergence 
is represented by the single pole in $\epsilon$. The divergence of double pole in the real- and virtual part is cancelled 
in the sum. The divergence in Eq.(\ref{DIV}) comes from the exchanged parton in the kinematic region of momentum which is collinear 
to the initial hadron $h$ or the final  $h'$. This divergent contribution from the collinear region is in fact already included in the tree-level 
result of the chiral-even part of our first observable in Eq.(\ref{TreeO}), i.e., in the twist-3 parton distributions
and twist-2 parton FF's. Therefore, this divergent contribution needs to be subtracted to avoid a
double counting.      

\par 
\par 
The subtraction can be done by the replacement in the chirality-even part of the tree-level results in Eq.(\ref{TreeO}): 
\begin{equation} 
  T_F(x,x) \to  T_F(x,x) - \Delta T_F(x,x), \quad \hat d(z)   \to  \hat d(z)  - \Delta\hat d(z). 
\end{equation}
The contributions which should be added in the final results for the subtraction are:
\begin{eqnarray}
\Delta \frac{ d \sigma \langle {\mathcal O}_1\rangle}{ d x_B dy d z_h}  &=& - \pi\frac{z_h y\alpha^2 \alpha_s}{ Q^2} 
  \vert s_\perp\vert^2 Y_M  \biggr (1-\frac{\epsilon}{2} \biggr ) \biggr [ \Delta\hat d(z_h) T_F(x_B,x_B) + 
     \hat d(z_h) \Delta T_F(x_B,x_B) \biggr],   
\nonumber\\
\Delta \frac{ d \sigma \langle {\mathcal O}_2\rangle}{ d x_B dy d z_h} &=& 0.  
\label{SUB} 
\end{eqnarray} 
In dimensional regularization for regularizing all divergences, the quantities $\Delta\hat d(z)$ and $\Delta T_F(x,x)$ 
are determined by the evolution kernels. The evolution kernel of the quark fragmentation function is well-known. 
The evolution of $T_F(x,x)$ is studied in \cite{KangQiu1, BMP, ZYL, MW, SchZh, KangQiu2}.
At one-loop level we have: 
\begin{eqnarray} 
\Delta T_F(x,x) &=& \frac{\alpha_s}{2\pi} \left ( -\frac{2}{\epsilon_c} + \ln \frac{e^\gamma \mu^2}{4\pi\mu_c^2 }\right ) 
 \biggr \{-N_c T_F(x,x) +   
    \int_x^1  \frac{dz}{z}   \biggr  [ {\mathcal P}_{qq} (z) T_F(\xi,\xi)  + \frac{N_c}{2} \biggr ( T_{\Delta}(x,\xi)  
\nonumber\\    
   &&  +  \frac{ (1+z) T_F(x,\xi) -(1+z^2) T_F(\xi,\xi) }{1-z} \biggr ) 
+ \frac{1}{2N_c} \biggr (  (1-2z) T_{F}(x,x-\xi)
\nonumber\\ 
    &&   + T_{\Delta} (x,x-\xi ) \biggr )  
  -\frac{1}{2}   \frac{ (1-z)^2 + z^2 }{\xi}    
     T_{G+}(\xi,\xi)  \biggr ]  \biggr \} 
\nonumber\\    
  &=&  \frac{\alpha_s}{2\pi} \left ( -\frac{2}{\epsilon_c} + \ln \frac{e^\gamma \mu^2}{4\pi\mu_c^2 }\right )
       \biggr ( {\mathcal F}_q \otimes T_F + {\mathcal F }_{\Delta q} \otimes T_\Delta + {\mathcal  F }_{g} \otimes T_{G+}    \biggr ) (x),  
\nonumber\\
  \Delta \hat d(x)  &=&  \frac{\alpha_s }{2\pi} \left ( -\frac{2}{\epsilon_c} + \ln \frac{e^\gamma \mu^2}{4\pi\mu_c^2 }\right ) 
     \int \frac{d\xi}{\xi} \biggr \{  P_{qq}(z  ) \hat d (\xi)
       +  P_{gq}(z) \hat g (\xi) \biggr \}  
\nonumber\\
    &=&  \frac{\alpha_s }{2\pi} \left ( -\frac{2}{\epsilon_c} + \ln \frac{e^\gamma \mu^2}{4\pi\mu_c^2 }\right ) 
      \biggr ( {\mathcal P}_{qq}\otimes \hat d + {\mathcal P} _{gq } \otimes \hat g \biggr ) (x),         
\label{SUBTFQ}     
\end{eqnarray}
with $z=x/\xi$. In the above we define five convolutions ${\mathcal F}$'s and ${\mathcal P}$'s  for short notations.
$\mu$ is the renormalization scale. $\mu_c$ is the scale related to collinear divergences.  
The derivative of $\Delta T_F(x,x)$ with respect to $\mu$ gives the evolution kernel of $T_F(x,x)$ derived. 
Adding the contributions in Eq.(\ref{SUB}) to our calculated one-loop 
correction, we find that the chirality-even part of the one-loop correction is finite. 
\par 
The chirality-even corrections to our first observable have been studied in \cite{KVX,DKPV,ShYo}. 
In \cite{DKPV} the contributions from Fig.10 are obtained, where there is a soft-gluon-pole contributions involving the derivative 
of $N(x,x)$ or $N(x,0)$ with respect to $x$, and the derivative of $O(x,x)$ or $O(x,0)$. These derivative terms can be 
eliminated by integration by part. It is noted that there is an integration over $P_{h\perp}$ in Eq.(\ref{IOS}) 
for our spin observables. With the master formulas in \cite{KoTa,KoTa2,KTY}, such derivative terms are eliminated automatically through the integration over $P_{h\perp}$. This also holds for the case of $T_F(x,x)$, where there is a derivative contribution of $T_F(x,x)$ 
in \cite{KVX}. In \cite{KVX} the contributions from Fig.6 and Fig.8 are obtained, but without the contributions from $T_{\Delta}(x_1,x_2)$. In \cite{ShYo} the contributions  with $T_{\Delta}(x_1,x_2)$ from Fig.6 and Fig.8 and 
the contributions from Fig.7 are included. But, the soft-quark-pole contributions are still missing. 
In all these works, only the first term with $g^{\mu\nu}$ in the first equation in Eq.(\ref{IPSI}) is taken into account
and the contributions from the second term with $Y_L$ are not considered.   
It is true that the second term with $Y_L$ does not contribute at tree-level. But it will contribute 
beyond tree-level. Except these missing contributions our results agree with those in \cite{KVX,DKPV,ShYo}. 
The missing contributions with $Y_L$ and the soft-quark-pole contributions are included in this work, they are finite.

\par\vskip20pt
\noindent 
{\bf 5. Final Results }  
\par 
As mentioned in previous sections, the evolutions of involved parton distributions and FF's take the forms of convolutions. 
We denote these evolutions as: 
\begin{eqnarray} 
 \frac{ \partial \hat d(z)}{\partial \ln \mu} &=& \frac{\alpha_s}{\pi} \biggr ( {\mathcal P}_{qq}\otimes \hat d + {\mathcal P} _{gq } \otimes \hat g \biggr ) (z), 
\nonumber\\
 \frac{ \partial T_F (x,x)}{\partial \ln \mu} &=& \frac{\alpha_s}{\pi}\biggr ( {\mathcal F}_q \otimes T_F + {\mathcal F }_{\Delta q} \otimes T_\Delta + {\mathcal  F }_{g} \otimes T_{G+}    \biggr ) (x), 
\nonumber\\
 \frac{  \partial h_1 (x)}{\partial \ln \mu} &=& \frac{\alpha_s}{\pi} \biggr ( P_{\perp q}\otimes h_1 \biggr ) (x), 
\nonumber\\
\frac{   \partial \hat e_\partial (z)}{\partial \ln \mu} &=& \frac{\alpha_s}{\pi} \biggr ( {\mathcal F}_\partial \otimes \hat e_\partial + {\mathcal F}_F \otimes \hat E_F + {\mathcal F}_G \otimes 
        \hat E_G \biggr ) (z). 
\end{eqnarray}         
The definitions of the convolutions can be found in Eq.(\ref{Dh1ep}) and Eq.(\ref{SUBTFQ}).  
With these notations we can write our final results in the form which are explicitly $\mu$-independent at one-loop level:
\begin{eqnarray} 
\frac{d \sigma \langle {\mathcal O}_1 \rangle }{ d x_B dy d z_h} &=& \frac{\pi z_h y \alpha^2}{Q^2} \vert s_\perp\vert^2  Y_M \biggr [   \hat d (z_h) T_F (x_B, x_B) -\frac{1 }{2} \ln\frac{e\mu^2}{Q^2}  \frac{\partial}{\partial \ln \mu}  \biggr (  
  T_F (x_B,x_B) \hat d(z_h) \biggr )    
\nonumber\\
  && - \frac{\partial}{\partial \ln\mu} \biggr (h_1 (x_B) \hat e_\partial (z_h)\biggr ) -\frac{\alpha_s}{2\pi}C_F
  \biggr ( 5  \hat d(z_h) T_F (x_B, x_B) - 6  h_1 (x_B) \hat e_\partial (z_h)\biggr )
\nonumber\\
   && + \frac{\alpha_s}{8\pi} \hat d(z_h) \int \frac{ dx }{x^2} ( (1-\hat x)^2 + \hat x^2 ) \biggr ( 3 T_{G+} (x,x)-2 T_{G-}(x,0) \biggr ) \biggr ] + \frac{d \sigma \langle {\mathcal O}_1 \rangle _F}{ d x_B dy d z_h}, 
\nonumber\\
\frac{d \sigma \langle {\mathcal O}_2 \rangle }{ d x_B dy d z_h} &=& 2 \pi z_h y \alpha^2 \vert s_\perp\vert^2 Y_2  
    \biggr [  2 h_1 (x_B)  \hat e_\partial (z_h) -\biggr ( \ln\frac{\mu^2}{Q^2} +\frac{3}{2} \biggr ) \frac{\partial }{\partial\ln\mu} 
    \biggr (h_1 (x_B)  \hat e_\partial (z_h) \biggr ) 
\nonumber\\    
    && -\frac{5 \alpha_s}{2\pi}  C_F h_1 (x_B) \hat e_\partial (z_h) \biggr ]  
    + \frac{d \sigma \langle {\mathcal O}_2 \rangle_F }{ d x_B dy d z_h}, 
\label{DIVP}                            
\end{eqnarray} 
where the last term with the sub-index $F$  in the result of our two observables is the sum of all finite contributions from sets of diagrams 
studied in previous sections. The sums are: 
\begin{eqnarray} 
\frac {d\sigma \langle {\mathcal O}_1\rangle _F }{ d x_B d y d z_h}  &=& \frac{z_h \alpha_s \alpha^2 y}{4 Q^2} \vert s_\perp\vert^2 
\int \frac{ d x d z}{x z}  \biggr \{   h_1 (x) \biggr [  
  \hat e_\partial (z)  {\mathcal A}_{1\sigma\partial} (\hat x, \hat z) 
\nonumber\\   
   && 
 + 2 \int\frac{ d z_1}{z_1} \biggr ( {\rm Im} \hat E_F (z_1,z) {\mathcal A}_{1\sigma F} (\hat x, \hat z,\hat z_1)
  + {\rm Im } \hat E_G (z_1,z) {\mathcal A}_{1\sigma G} (\hat x, \hat z) \biggr ) \biggr ] 
\nonumber\\
   && +  \hat d (z) \biggr [ 
 T_F (x,x_B) {\mathcal A}_{1hq} (\hat x, \hat z) 
      + T_\Delta  (x,x_B) {\mathcal B}_{1hq} (\hat x, \hat z) +T_F (x_B-x,x_B) {\mathcal C}_{1Fq} (\hat x,\hat z)  
\nonumber\\
     &&+ T_{\Delta} (x_B-x,x_B) {\mathcal C}_{1Dq} (\hat x, \hat z)         + T_F (x,x) {\mathcal D}_{1q} (\hat x,\hat z)  + T_F(-x,0) {\mathcal E}_{1Fq} (\hat x,\hat z)
\nonumber\\
     && +T_\Delta (-x,0) {\mathcal E}_{1\Delta q} (\hat x,\hat z) +  \frac{1}{x} T_{G+} (x,x)  {\mathcal F}_{1g+}(\hat x,\hat z) 
      + \frac{1}{x} T_{G-} (x,0)  {\mathcal F}_{1g-}(\hat x,\hat z)   \biggr ] 
\nonumber\\
   && -\hat d(-z) \biggr [  T_F (x_B-x,x_B) {\mathcal C}_{1F\bar q} (\hat x,\hat z)  
  + T_{\Delta} (x_B-x,x_B) {\mathcal C}_{1D\bar q} (\hat x, \hat z) 
\nonumber\\  
  && + T_F(0,x) {\mathcal E}_{1F\bar q} (\hat x,\hat z) 
 +T_\Delta (0, x) {\mathcal E}_{1\Delta\bar  q} (\hat x,\hat z)  \biggr ]     
\nonumber\\     
     &&+ \hat g (z)   \biggr [  
 T_F (x,x_B) {\mathcal A}_{1hg} (\hat x, \hat z)  
      + T_\Delta  (x,x_B) {\mathcal B}_{1hg} (\hat x, \hat z) 
\nonumber\\      
     &&   + T_F (x,x) {\mathcal D}_{1g}(\hat x,\hat z) 
      + T_F(x,0) {\mathcal E}_{1Fg} (\hat x,\hat z)         
 +T_\Delta (x,0) {\mathcal E}_{1\Delta g} (\hat x,\hat z)\biggr ]  \biggr \},
\nonumber\\
   && 
\nonumber\\
  \frac {d\sigma \langle {\mathcal O}_2\rangle _F }{ d x_B d y d z_h}  &=& \frac{z_h \alpha_s \alpha^2 y}{4 } \vert s_\perp\vert^2 
\int \frac{ d x d z}{x z}  \biggr \{   h_1 (x) \biggr [  
  \hat e_\partial (z)  {\mathcal A}_{2\sigma\partial} (\hat x, \hat z) 
\nonumber\\   
   && 
 + 2 \int\frac{ d z_1}{z_1} \biggr ( {\rm Im} \hat E_F (z_1,z) {\mathcal A}_{2\sigma F} (\hat x, \hat z,\hat z_1)
  + {\rm Im } \hat E_G (z_1,z) {\mathcal A}_{2\sigma G} (\hat x, \hat z) \biggr ) \biggr ] 
\nonumber\\
  && +  \hat d (z)  \biggr [  
    T_F (x_B-x,x_B) {\mathcal C}_{2Fq} (\hat x,\hat z)+ T_{\Delta} (x_B-x,x_B) {\mathcal C}_{2Dq} (\hat x, \hat z)   
\nonumber\\
     &&        + T_F(-x,0) {\mathcal E}_{2Fq} (\hat x,\hat z)
   + T_\Delta (-x,0) {\mathcal E}_{2\Delta q} (\hat x,\hat z) + \frac{2 \hat x^2}{x \hat z} ( T_{G+} (x,x) + 2 T_{G-}(x,0) )  \biggr ] 
\nonumber\\
   &&  - \hat d(-z) \biggr [  T_F (x_B-x,x_B) {\mathcal C}_{1F\bar q} (\hat x,\hat z)  
  + T_{\Delta} (x_B-x,x_B) {\mathcal C}_{1D\bar q} (\hat x, \hat z) 
\nonumber\\  
  && + T_F(0,x) {\mathcal E}_{1F\bar q} (\hat x,\hat z) 
 +T_\Delta (0,x) {\mathcal E}_{1\Delta\bar  q} (\hat x,\hat z)  \biggr ]       
\nonumber\\
   && + \hat g(z) \biggr [ T_F(x,0) {\mathcal E}_{2Fg} (\hat x,\hat z) +  T_\Delta (x,0) {\mathcal E}_{2\Delta g} (\hat x,\hat z)  \biggr ]\biggr \}.
\label{FINP}     
\end{eqnarray} 
Eq.(\ref{DIVP},\ref{FINP}) are our main results. The perturbative functions ${\mathcal A}$'s, ${\mathcal B}$'s, ${\mathcal C}$'s and ${\mathcal E}$'s  are given in Appendix. 
In Eq.(\ref{TreeO}) and the results in this section the contributions from charge-conjugated parton processes are not included. They can be obtained from our results with charge-conjugation.

\par\vskip40pt
\noindent 
{\bf 6.  Summary} 
\par 
In this work we have studied two spin observables of SIDIS in which the initial hadron is transversely polarized. 
They are weighted differential cross-sections corresponding to Sivers- or Collins asymmetry, respectively.  
These asymmetries have been measured in experiment already.  In fact, one of the studied observables is weighted Sivers asymmetry, while another is weighted Collins asymmetry. 
In collinear factorization 
they take factorized forms as convolutions of perturbative coefficient functions with twist-3 parton distributions 
combined twist-2 FF's or twist-2 transversity distribution combined with twist-3 FF's. The perturbative coefficient functions
have been calculated at one-loop level in this work. The collinear divergences are correctly subtracted so that these functions 
are finite. With our results the spin observables or 
SSA's are predicted more precisely than with tree-level results. The spin observables 
studied here can already be measured in current COMPASS- and JLab experiment, and in future experiment at EIC. 

\par 

It is interesting to note that at tree-level our first observable is predicted only with ETQS matrix element and the twist-2 quark FF, and the second observable is predicted only with the twist-2 transversity distribution and one  of twist-3 quark FF's.
This implies that through the measurement of these observables, it is possible to determine these nonperturbative 
quantities at tree-level accuracy.  But, at one-loop level more twist-3 parton distributions and FF's are involved.
To determine all involved parton distributions and FF's, one has to combine theoretical- and experimental results from other processes,  like Drell-Yan 
processes and inclusive single hadron production at hadron-hadron collisions, where the involved parton distributions and FF's also appear in theoretical predictions.  This requires more studies both in theory and experiment.

\par\vskip20pt
\noindent
{\bf Acknowledgments}
\par
The work is supported by National Nature
Science Foundation of P.R. China(No.11275244, 11675241, 11605195). The partial support from the CAS center for excellence in particle 
physics(CCEPP) is acknowledged.

\par\vskip20pt
\renewcommand{\theequation}{A.\arabic{equation}}
\setcounter{equation}{0}
\par 

\par\vskip30pt
\noindent
{\bf Appendix}

\par
We list here all functions appearing in the finite part of our results. We define two functions
\begin{eqnarray} 
   L_{\pm}(\xi) = \left ( \frac{\ln(1-\xi)}{1-\xi}\right )_+ \pm \frac{\ln\xi}{1-\xi}. 
\end{eqnarray}
The functions in the chirality-odd contributions in Eq.(\ref{1LRCO}) are:     
\begin{eqnarray}
\mathcal {A}_{1 \sigma \partial} &=& 8Y_M C_F \biggr ( \delta(1-\hat{x})
\frac{1}{(1-\hat{z})_+}
 + \delta(1-\hat{z}) \frac{\hat{x}}{(1-\hat{x})_+} \biggr ),
\\
\mathcal {A}_{1\sigma F} &=&
 \frac{Y_M}{\hat{z} (\hat{z}-\hat{z}_1)^2} \biggr [
2C_F \hat{z}_1 \biggr ( - 3\hat{z}^2 +
\hat{z}(\hat{z}_1 + 4) - 2 \hat{z}_1 \biggr )
 - \frac{1}{N_c} \hat{z} \biggr ( \hat{x} (\hat{z}_1-\hat{z}) +
2(\hat{z}-1) \hat{z}_1 \biggr ) 
\nonumber\\
   && - \delta (1-\hat{x})
\frac{\hat{z}^2}{ \hat{z}_1 - 1}
 \biggr ( 2 C_F \hat{z}_1 ( \hat{z}^2 - 2\hat{z} \hat{z}_1 + \hat{z} + \hat{z}_1^2 + \hat{z}_1-2 ) -\frac{1}{N_c} (
\hat{z}+\hat{z}_1-\hat{z}^2-\hat{z}_1^2) \biggr ) \biggr ]
\nonumber\\ &&
- \frac{ Q^2 Y_L \hat{x} \hat{z}}{2 x_B^2   (\hat{z} - \hat{z}_1)
(\hat{x} \hat{z} - \hat{x} \hat{z}_1-\hat{z} \hat{z}_1 + \hat{z}_1)}
\biggr [ 2 C_F \hat{z} \hat{z}_1 (\hat{z}_1-\hat{z})
   - \frac{1}{N_c}(\hat{z}-1) (\hat{x}\hat{z}-\hat{x}
\hat{z}_1 + \hat{z}_1) \biggr ],
\\
\mathcal {A}_{1\sigma G} &=&  -Y_M \delta(1-\hat{x})
\frac{2 C_F (\hat{z}-1)^2}{N_c (\hat{z}_1 + 1 - \hat{z})} -
\biggr ( Y_M + \frac{Q^2 Y_L (\hat{z} - 1) \hat{z}}{2 x_B^2} \biggr ) \frac{ 2 C_F\hat{x}(\hat{z} - 1) } {\hat{z}(\hat{z} - \hat{z}_1)
(\hat{x} \hat{z} - \hat{x} \hat{z}_1 - \hat{z} \hat{z}_1 +
\hat{z}_1)}
\nonumber\\
&& \times \frac{ \hat{x}^2 \hat{z}_1^2 + \hat{x} (\hat{z}^3 -
3\hat{z}^2 \hat{z}_1 + 2\hat{z} \hat{z}_1^2 + \hat{z} \hat{z}_1 -
2\hat{z}_1^2 ) + (\hat{z}-1)^2 \hat{z}_1 (\hat{z}_1 - \hat{z})} {
N_c ( \hat{x} \hat{z}_1 - \hat{z}^2 +
\hat{z} \hat{z}_1 + \hat{z} - \hat{z}_1 ) }, 
\\
\mathcal {A}_{2\sigma\partial} &=& 8 C_F \biggr \{ -\delta ( 1 - \hat{x} ) \delta (1-\hat{z}) + \delta(1-\hat{x})
\biggr ( L_+(\hat{z})  + \frac{3 }{ (1-\hat{z})_+} \biggr ) 
\nonumber\\
&& + \hat{x } \delta(1-\hat{z}) \biggr [ 2 L_-(\hat{x})  +
\frac{3}{(1-\hat{x})_+} \biggr ] + \frac{2 \hat{x} }
{(1-\hat{x})_+(1-\hat{z})_+} \biggr \},
\\
\mathcal {A}_{2\sigma F} &=& \frac{4}{\hat{z}-\hat{z}_1}\delta(1-\hat{x}) \biggr [- L_+(\hat{z})(1-\hat{z}) \hat{z} \biggr ( C_F
(\hat{z}-\hat{z}_1-1) - \frac{N_c}{2} \frac{\hat{z}^2 + \hat{z}_1^2 - \hat{z} - \hat{z}_1 } {(\hat{z} - \hat{z}_1) (1-\hat{z}_1)} \biggr )
\nonumber\\ && - \frac{1}{\hat{z}} \biggr ( \frac{C_F}{2}  ( 3\hat{z}^3 -
3\hat{z}^2 (\hat{z}_1 + 1) + 2\hat{z}(\hat{z}_1+1) - 2\hat{z}_1  )
 - \frac{N_c}{4} \frac{ 1 }{(\hat{z}-\hat{z}_1) (1-\hat{z}_1)}
\nonumber\\
&& \times \hat{z}^2 (3\hat{z}^2 - 2\hat{z} \hat{z}_1^2 - 3\hat{z} + 5\hat{z}_1^2 - 5\hat{z}_1+2) \biggr ) \biggr ] +  \frac{ 4\hat{x} \hat{z}} {(1-\hat{x})_+ (\hat{z} - \hat{z}_1 )^2}
\nonumber\\
&& \times \frac{ 1 } {  (\hat{x} \hat{z} - \hat{x}\hat{z}_1 - \hat{z}\hat{z}_1 + \hat{z}_1)} \biggr [\frac{N_c^2-1}{2N_c} \hat{z}_1 \biggr (\hat{x}\hat{z}^3
 + \hat{x}\hat{z}^2 (1-2\hat{z}_1) + \hat{x} \hat{z} (\hat{z}_1^2 - 2\hat{z}_1 - 2)
\nonumber\\
&& + \hat{x}\hat{z}_1(\hat{z}_1+2) + \hat{z}_1(3\hat{z} - \hat{z}_1-2) \biggr ) + \frac{1}{2N_c} \biggr (\hat{x}^2 (\hat{z}-1)(\hat{z}-\hat{z}_1)^2
\nonumber\\
&&  + \hat{x} (2\hat{z} - 3) \hat{z}_1 (\hat{z} -
\hat{z}_1 ) + 2(\hat{z}-1) \hat{z}_1^2 \biggr )\biggr ],
\\
\mathcal {A}_{2\sigma G} &=& 2 C_F \delta(1-\hat{x}) \biggr [  L_+(\hat{z})(1 - \hat{z})(2\hat{z}-2) \hat{z}  + (3\hat{z}^2 - 3\hat{z} + 2\hat{z}_1 + 2 )
\bigg ]
 \frac{ (1-\hat{z})  }{ N_c\hat{z} (1+\hat{z}_1 - \hat{z}) }
\nonumber\\ 
  && - \frac{ 4 C_F \hat{x}^2 (\hat{z}-1)^2 }{(1-\hat{x})_+  (\hat{z} - \hat{z}_1) (\hat{x} \hat{z} - \hat{x} \hat{z}_1 - \hat{z} \hat{z}_1 + \hat{z}_1 ) }
\nonumber\\
&& \times \frac{ \hat{x}^2 \hat{z}_1^2 + \hat{x} (\hat{z}^3 -
3\hat{z}^2 \hat{z}_1 + 2\hat{z} \hat{z}_1^2 + \hat{z} \hat{z}_1 -
2\hat{z}_1^2 ) + (\hat{z} - 1)^2 \hat{z}_1 (\hat{z}_1 - \hat{z})}
{ N_c (\hat{x} \hat{z}_1 - \hat{z}^2 + \hat{z}\hat{z}_1 + \hat{z} - \hat{z}_1) }.
\end{eqnarray}
The functions in the chirality-even contributions in Eq.(\ref{HP1},\ref{HP2},\ref{HSG},\ref{HSQ},\ref{H3G}) are:
\begin{eqnarray}
 \mathcal {A}_{1 h q} (\hat{x}, \hat{z}) &=& Y_M \biggr [ \frac{1 + \hat{z}
(N_c^2-1) }{ \hat{z} N_c} \delta(1  - \hat{x})
 \biggr ( L_+(\hat{z})(1 + \hat{z}^2) - \hat{z} + 1 \biggr )
+ N_c\delta(1-\hat{z})
\nonumber\\
&& \times  L_-(\hat{x}) (1 + \hat{x})
 + \frac{1 + \hat{z}(N_c^2-1)}{\hat{z} N_c}
\frac{1 + \hat{x} \hat{z}^2}
 {(1 - \hat{x})_+(1-\hat{z})_+} \biggr ],
 \\
\mathcal {B}_{1 h q} (\hat{x}, \hat{z}) &=& Y_M \biggr [ N_c\delta(1-\hat{z})
L_-(\hat{x})( \hat{x}-1 )
 + \frac{1 + \hat{z}(N_c^2-1)}{\hat{z}N_c}
\frac{\hat{x}\hat{z}^2-1}{(1-\hat{x})_+(1-\hat{z})_+} \biggr ],
\\
\mathcal {A}_{1hg} (\hat{x}, \hat{z}) &=& Y_M \biggr [ \delta(1-\hat{x})
 \biggr ( - L_+(\hat{z})(1 - \hat{z})(2 -
2\hat{z}+\hat{z}^2) - \hat{z}^2 \biggr )
 -\frac{1+\hat{x}(1-\hat{z})^2}{(1-\hat{x})_+} \biggr ]
 \nonumber\\
 && \times \frac{ N_c^2 (1-\hat{z}) + \hat{z}}{\hat{z}^2 N_c},
 \\
 \mathcal {B}_{1hg}
(\hat{x}, \hat{z}) &=& Y_M \frac{ N_c^2 (1-\hat{z}) + \hat{z} }{\hat{z}^2 N_c}
\frac{\hat{x}(1-\hat{z})^2-1}{\hat{x}-1}, 
\\
\mathcal {C}_{1F q} (\hat{x}, \hat{z}) &=&  Y_M \biggr [\frac{1}{N_c} \biggr (
\delta(1-\hat{z}) L_-(\hat{x}) (1-\hat{x}) (1-2\hat{x})   +
\frac{(1-2\hat{x}) \hat{z}}{(1-\hat{z})_+} \biggr)
\nonumber\\
&& - \frac{1}{\hat{z} }(2\hat{x}-1) (2\hat{z}^2-2\hat{z}+1) \biggr ],
\\
\mathcal {C}_{1D q} (\hat{x}, \hat{z}) &=& -Y_M \biggr [ \frac{1}{N_c} \biggr
( \delta(1-\hat{z}) L_-(\hat{x})(1-\hat{x}) + \frac{\hat{z}}{(1-\hat{z})_+} \biggr ) +\frac{2\hat{z}^2 - 2\hat{z} + 1}{\hat{z} } \biggr ], 
\\
\mathcal {C}_{1F \bar q} (\hat{x}, \hat{z}) &=&  -Y_M \biggr [\frac{(2\hat{x} -1 )(\hat{z}-1)^2}{N_c \hat{z}^2} + \frac{(2\hat{x}-1)
(2\hat{z}^2 - 2\hat{z} + 1)}{\hat{z} } \biggr ],
\\
\mathcal {C}_{1D \bar q} (\hat{x}, \hat{z})  &=&  - Y_M \biggr [ \frac{(\hat{z}-1)^2}{N_c \hat{z}^2} + \frac{ 2\hat{z}^2 - 2\hat{z} + 1}{\hat{z} } \biggr ],
\\
\mathcal {C}_{2F q} (\hat{x}, \hat{z}) &=&   -\frac{1}{N_c} 2 \hat x +
4 \hat x (\hat z -1),
\\
\mathcal {C}_{2D q} (\hat{x}, \hat{z}) &=&  \frac{1}{N_c} 2
\hat x - 4 \hat x (\hat z -1),
\\
\mathcal {C}_{2F \bar q} (\hat{x}, \hat{z}) &=& \frac{1}{N_c} 2 \hat x
\frac{\hat z -1}{\hat z} + 4 \hat x (\hat z -1),
\\
\mathcal {C}_{2D \bar q} (\hat{x}, \hat{z}) &=& -\frac{1}{N_c} 2 \hat
x \frac{\hat z -1}{\hat z} - 4 \hat x (\hat z -1), 
\\
\mathcal {D}_{1q} (\hat{x}, \hat{z}) &=& \frac{-1}{\hat{z} N_c}\biggr
\{ Y_M \biggr [\delta(1-\hat{x})  \biggr (  L_+(\hat{z})(1 +
\hat{z}^2) +
\frac{2(\hat{z}^2 - \hat{z} + 1)}{(1 - \hat{z})_+} \biggr ) \nonumber\\
&& + \delta(1 - \hat{z}) \biggr (  L_-(\hat{x})(1 + \hat{x}^2) +
\frac{2(\hat{x}^2 - \hat{x}+1)}{(1 - \hat{x})_+}  \biggr ) \nonumber\\
&& + \frac{(1 - \hat{x})^2 + (1-\hat{z})^2 + 2\hat{x} \hat{z}}
{(1 - \hat{x})_+ (1 - \hat{z})_+} \biggr ] + \frac{Y_L Q^2}{ 2x_B^2}
\hat{x} \hat{z}  \biggr \},
\\
\mathcal {D}_{1g} (\hat{x}, \hat{z}) &=& \frac{N_c}{\hat{z}^2} \biggr \{ Y_M \biggr [
\delta(1-\hat{x}) L_+(\hat{z}) ( 1 - \hat{z} )( \hat{z}^2 + 2 -
2\hat{z} )  + \delta(1 - \hat{x})
\nonumber\\
&& \times ( 2 \hat{z}^2 + 2-2 \hat{z} )
 + \frac{(1-\hat{x})^2
 + \hat{z}^2 + 2\hat{x}(1 - \hat{z})}{(1 - \hat{x})_+} \biggr ]
 + \frac{Y_LQ^2}{2x_B^2} \hat{x}(1 - \hat{z}) \hat{z} \biggr \}, 
\\
 \mathcal {E}_{1Fq} (\hat{x}, \hat{z}) &=& \frac{1}{\hat{z}N_c} \biggr (  \frac{Y_M}{\hat{z}} (-2
\hat{x}^2 + 2 \hat{x} \hat{z} + \hat{x} -\hat{z}^2 ) - \frac{ Y_L
Q^2 }{x_B^2 } \hat{x} (\hat{x}-1) (\hat{z}-1) \biggr ),
\\
\mathcal {E}_{1 \Delta q} (\hat{x}, \hat{z}) &=&  \frac{Y_M}{\hat{z}^2
N_c} (\hat{x} + \hat{z}^2 - 2 \hat{x} \hat{z} ),
\\
\mathcal {E}_{1F \bar q } (\hat{x}, \hat{z}) &=& \frac{1}{\hat{z}N_c} \biggr (  \frac{Y_M}{\hat{z}} (2
\hat{x}^2 - 2 \hat{x} \hat{z} - \hat{x} + \hat{z}^2 ) + \frac{ Y_L
Q^2 }{x_B^2 } \hat{x} (\hat{x}-1) (\hat{z}-1) \biggr ),
\\
\mathcal {E}_{1 \Delta \bar q } (\hat{x}, \hat{z}) &=& \frac{Y_M}{\hat{z}^2
N_c} (  \hat{x} + \hat{z}^2  - 2 \hat{x} \hat{z} ),
\\
\mathcal {E}_{1F g} (\hat{x}, \hat{z}) &=& - \frac{1}{\hat{z}N_c} \biggr (  \frac{Y_M}{\hat{z}} (-2 \hat{x}^2 + 2 \hat{x} \hat{z} + \hat{x} -\hat{z}^2 ) - \frac{ Y_L Q^2 }{x_B^2 } \hat{x} (\hat{x}-1) (\hat{z}-1) \biggr ),
\\
\mathcal {E}_{1 \Delta g} (\hat{x}, \hat{z}) &=&  - \frac{Y_M}{\hat{z}^2
N_c} (\hat{x} + \hat{z}^2 - 2 \hat{x} \hat{z} ),
\\
 \mathcal {E}_{2Fq} (\hat{x}, \hat{z}) &=& \frac{1}{\hat z N_c} 2\hat
x (\hat z -1)(2\hat x - 1),
\\
\mathcal {E}_{2 \Delta q} (\hat{x}, \hat{z}) &=&  \frac{1}{\hat z N_c}
2\hat x (\hat z -1),
\\
\mathcal {E}_{2F \bar q } (\hat{x}, \hat{z}) &=& - \frac{1}{\hat z N_c} 2\hat
x (\hat z -1)(2\hat x - 1),
\\
\mathcal {E}_{2 \Delta \bar q } (\hat{x}, \hat{z}) &=& \frac{1}{\hat z N_c}
2\hat x (\hat z -1),
\\
\mathcal {E}_{2F g} (\hat{x}, \hat{z}) &=& - \frac{1}{\hat z N_c} 2\hat
x (\hat z -1)(2\hat x - 1),
\\
\mathcal {E}_{2 \Delta g} (\hat{x}, \hat{z}) &=&  - \frac{1}{\hat z N_c}
2\hat x (\hat z -1), 
\\
\mathcal {F}_{1M+}
(\hat{x}, \hat{z}) &=&  \biggr ( L_-(\hat{x})
(1-\hat{x}) +1 \biggr )\delta(1-\hat{z}) (1-2\hat{x}
 + 2\hat{x}^2)  + \delta(1-\hat{z})
 \nonumber\\
&& + \frac{1-2\hat{x}-2 \hat{z} + 2 \hat{x}^2 + 2 \hat{z}^2}
{\hat{z}^2 (1-\hat{z})_+} ,
\\
\mathcal {F}_{1M-} (\hat{x}, \hat{z}) &=&  L_-(\hat{x})
(1-\hat{x}) \delta(1-\hat{z})  (1-2\hat{x}
 + 2\hat{x}^2)  + \delta(1-\hat{z})
 \nonumber\\
&& + \frac{1-2\hat{x}-2 \hat{z} + 2 \hat{x}^2 + 2 \hat{z}^2}
{\hat{z}^2 (1-\hat{z})_+} ,
\\
\mathcal {F}_{1L} (\hat{x}, \hat{z}) &=&  \frac{ Q^2}{x_B^2}
\frac{(1-\hat{x}) \hat{x}}{\hat{z}} .
\end{eqnarray}
The perturbative functions ${\mathcal F}_{1g+}$ and ${\mathcal F}_{1g-}$ in Eq.(\ref{FINP}) for the gluonic contributions 
are determined by ${\mathcal F}_{1M\pm}$ and ${\mathcal F}_{1L}$ as:
\begin{eqnarray} 
{\mathcal F}_{1g+} (\hat x,\hat z) &=& \frac{1}{2}Y_M \biggr ( {\mathcal F}_{1M-} (\hat x, \hat z) -3 {\mathcal F}_{1M+} (\hat x, \hat z) \biggr ) - Y_L {\mathcal F}_{1L} (\hat x, \hat z)
\nonumber\\
 {\mathcal F}_{1g-} (\hat x,\hat z) &=& Y_M {\mathcal F}_{1M-} (\hat x, \hat z) - Y_M {\mathcal F}_{1M+} (\hat x, \hat z). 
\end{eqnarray}

\par

\end{document}